\documentclass[preprint,showpacs,preprintnumbers,amsmath,amssymb]{revtex4}
\usepackage{graphicx}
\usepackage{dcolumn}
\usepackage{bm}
\newcommand{\be}{\begin{eqnarray}}
\newcommand{\ee}{\end{eqnarray}}

\begin{document}
\title{Wave propagation through Cantor-set media: 
  Chaos, scaling, and fractal structures}
\author{Kenta Esaki, Masatoshi Sato, and Mahito Kohmoto}
\affiliation{%
Institute for Solid State Physics,
Kashiwanoha 5-1-5, Kashiwa, Chiba, 277-8581, Japan
}%
\date{\today}
\begin{abstract}
Propagation of waves through Cantor-set media is investigated by
renormalization-group analysis.  
For specific values of wave numbers, transmission coefficients are shown to be 
governed by the logistic map, and in the chaotic region, they show
sensitive dependence on small changes of parameters of the system such
as the index of refraction.
For other values of wave numbers, 
our numerical results suggest that light transmits completely or
reflects completely by the Cantor-set media ${\rm C}_{\infty}$.
It is also shown that transmission coefficients exhibit a local scaling behavior near complete transmission if the complete transmission is achieved at a wave number $\kappa=\kappa^*$ with a rational $\kappa^*/\pi$. The scaling function is obtained analytically by using the Euler's totient function, and the local scaling behavior is confirmed numerically. 
\end{abstract}


\maketitle

\section{Introduction}

Wave properties in fractal \cite{ko90,xs91,mp94,mp94_2,av02,fc02,nh05,nh06,us05,ss04,ss04_2} 
and quasi-periodic \cite{MK83,KO84,MK86,mk87,wg94,th94}
structures in one dimension
have been of theoretical and practical interest over the past
two decades.
They are typical examples of self-similar structures,
and physical properties peculiar to them
have been explored.
In order to observe effects of quasiperiodicity,
optical experiments using dielectric multilayers 
of ${\rm SiO_2}$ and ${\rm TiO_2}$ films  
were performed for the Fibonacci multilayer \cite{wg94,th94}. 
In these experiments, scaling behaviors of the transmission
coefficients were observed, which had been predicted by the
renormalization-group theory \cite{mk87}. 
For fractal structures, optical wave propagation on Cantor multilayers has been studied
 by several authors \cite{xs91,mp94,mp94_2,av02,fc02,nh05,nh06,us05}. 
Self-similar structures of the transmission (or reflection) coefficients 
were obtained numerically \cite{xs91,mp94,mp94_2,av02,fc02,nh05}.
Moreover, resonant states of light were studied
\cite{nh06,us05}, which was
motivated by an experiment using a three-dimensional fractal cavity
called the Menger sponge \cite{mw04}.

In this paper, we reexamine the optical wave propagation 
through Cantor sequences on the basis of the renormalization-group
theory.
In addition to the self-similar structures mentioned above, we find interesting
chaotic behaviors of the transmission coefficients analytically.
It is well known that non-linear dynamical systems showing
chaotic behaviors often have strange attractors 
with fractal structures\cite{attractor},
but we show that the reverse is also possible.
Namely, chaotic dynamics is obtained from a fractal structure. 
In the following, we find that, for specific values of wave
numbers, the transmission coefficients show chaotic behaviors governed
by the logistic map. 
For these wave numbers, the transmission coefficients 
are very sensitive to small changes of parameters 
of the system such as the index of refraction.
This exotic behavior leads to rapid oscillations of the
transmission coefficients as functions of the index of refraction, 
which could be observed in an optical experiment.
For other wave numbers, our numerical study suggests that 
light eventually transmits completely or reflects completely for the
infinite generation of the Cantor sequences ${\rm C}_{\infty}$.

We will also find intriguing local scaling behaviors of the transmission
coefficients near complete transmission, which are distinct from the
self-similar structures found in Refs.\cite{xs91,mp94,mp94_2}.
The complete transmission can be regarded as a fixed point of the
renormalization-group equation, and the local scaling behaviors are obtained
on the basis of the renormalization-group theory.
It will be shown that if complete
transmission is achieved at a wave number $\kappa=\kappa^*$ with a rational 
$\kappa^*/\pi$, the
transmission coefficient around the wave number exhibits a local
scaling behavior.
The scaling relation is determined by the
rational number $\kappa^*/\pi$ 
through the Euler's totient function, and an
analytic expression of the scaling function will be presented.
We will also compare the analytical results with numerical data.

This paper is organized as follows.
In Sec.\ref{sec:formulation}, 
we briefly explain the Cantor sequences and the Cantor-set media.
We then formulate our problem
in terms of renormalization-group transformation.
In Sec.\ref{sec:results},
we classify wave propagation and find chaotic behaviors of transmission
coefficients governed by the logistic map for specific wave numbers.
Optical experiments to observe the
chaotic behaviors are also proposed.
For other wave numbers, complete transmission and 
complete reflection of light 
in the Cantor-set media C$_{\infty}$ are numerically suggested.
In Sec.\ref{sec:results2}, scaling behaviors of the transmission coefficients 
near complete transmission are 
analyzed on the basis of the renormalization-group equation.
Finally, in Sec.\ref{sec:summary}, we summarize our results and 
discuss a generalization of our analysis for generalized Cantor-set media.

\section{Transfer-matrix method and renormalization-group transformation}
\label{sec:formulation}

In this section, we make a formulation of our problem.
The Cantor sequences and the Cantor-set media are constructed,
and the transfer matrix method is introduced to study
wave propagation through them.
Then the renormalization-group equation is defined in terms of
the transfer matrix method.

\begin{figure}
 \begin{center}
  \includegraphics[width=7cm]{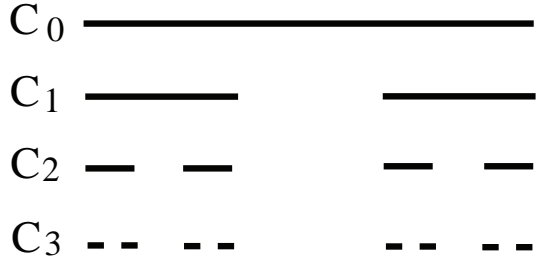}
\caption{\label{fig:cantorsequence}The first
  four generations of the Cantor sequences C$_j$.
The Cantor-set media C$_{\infty}$ are obtained in the $j\to \infty$ limit.}
\end{center}
\end{figure}

Let us construct the Cantor-set media first.
See Fig.\ref{fig:cantorsequence}.
The procedure of constructing the Cantor set begins with a line segment
with unit length (C$_0$ in Fig.\ref{fig:cantorsequence}). 
We regard this as substrate A. 
To obtain the first generation C$_1$, the line segment is
divided into three parts. The left and the right segments are
substrate A, each of which has length ${1}/{3}$, 
and the middle part, which has length ${1}/{3}$, is removed.
We regard the removed part as substrate B.
Then the procedure is repeated for each of remaining line segments A to
obtain new generations. 
We call the $j$th generation of the Cantor sequence as C$_j$.
For C$_j$ we have a set
of $2^j$ line segments of substrate A, each of which has length ${1}/{3^j}$.
By repeating this procedure infinite times,
we finally obtain the Cantor-set media, C$_{\infty}$, which 
are self-similar and have fractal dimension
$\log{2} /\log{3}$. 
In the following, we denote the indices of refraction of A and B 
as $n_{\rm A}$ and $n_{\rm B}$, respectively, and take $n_{\rm B}=1$
without loss of generality.

\begin{figure}
 \begin{center}
  \includegraphics[width=7.0cm,clip]{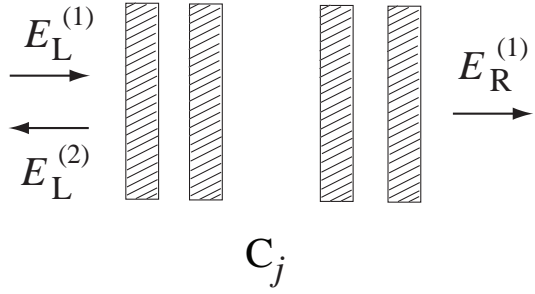}
\caption{\label{fig:setup}
  Electromagnetic wave propagation through the Cantor sequence C$_j$ ($j=2$).}
\end{center}
\end{figure}

\begin{figure}
 \begin{center}
  \includegraphics[width=8.0cm,clip]{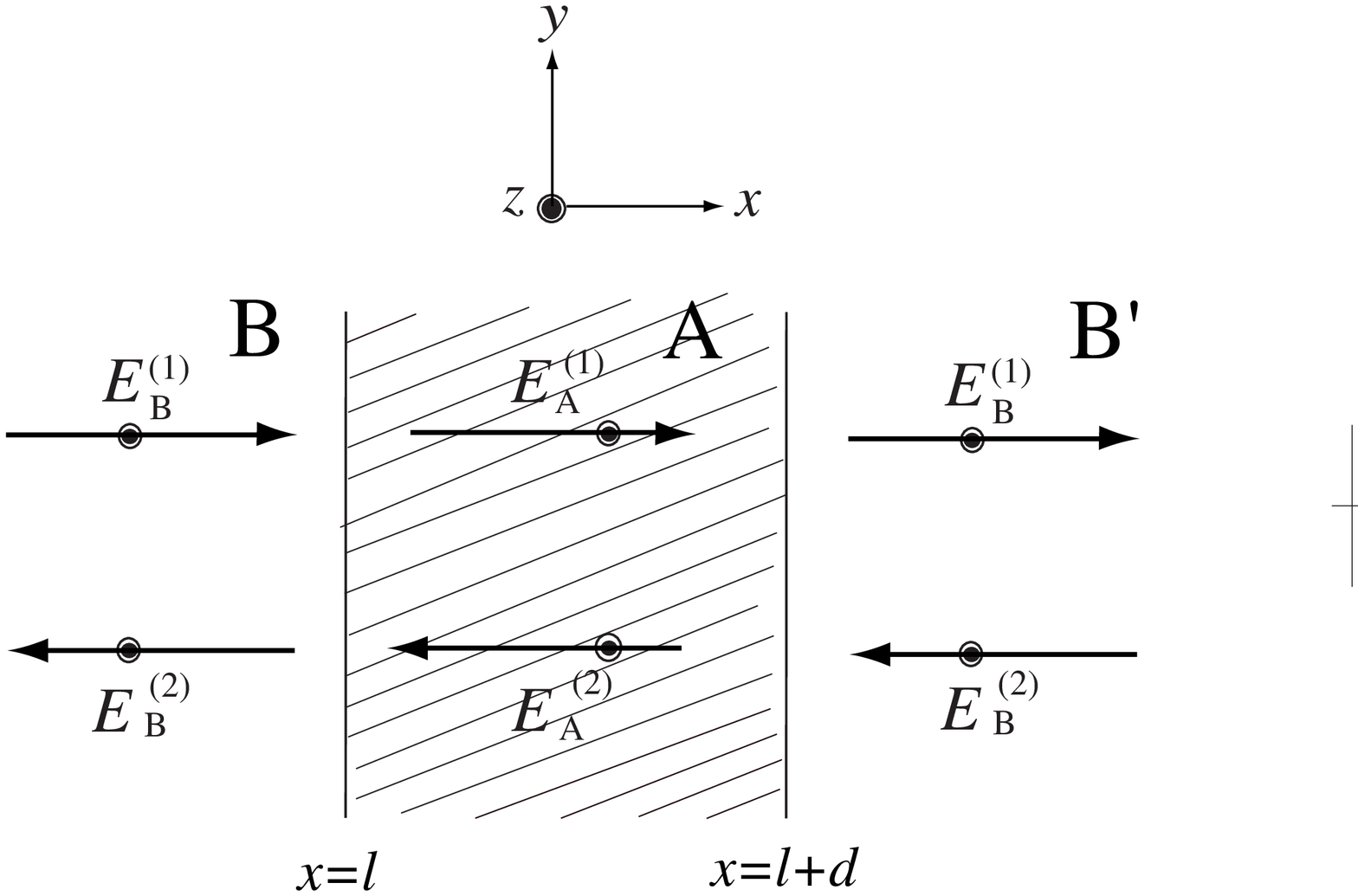}
\caption{\label{fig:interface}Electromagnetic wave propagation across
  interfaces between two layers A and B.} 
\end{center}
\end{figure}

Now consider wave propagation through the Cantor sequence ${\rm C}_j$ 
illustrated in Fig.\ref{fig:setup}.
For simplicity, we suppose that the incident light is linearly polarized.
Here $E_{\rm L}^{(1)}$, $ E_{\rm L}^{(2)}$ and $E_{\rm R}^{(1)}$ denote
the incident light, the reflected light and the transmitted light,
respectively.
No incoming wave from the right $E_{\rm R}^{(2)}$
exists, $E_{\rm R}^{(2)}=0$.
In order to understand light propagation through C$_j$,
let us first consider interfaces of two layers in Fig.\ref{fig:interface}. 
 The electric field for light in
layer A is given by
\begin{equation}
E= 
E_{\rm A}^{(1)}\exp[i k_{\rm A} x-i \omega t]
+ E_{\rm A}^{(2)}\exp[- i k_{\rm A} x -i \omega t],
\end{equation}
where $k_{\rm A}=n_{\rm A} k$ is the wave number of light in substrate A
and $\omega$ is frequency of light.
($k$ is a wave number of light in the vacuum.)
The electric field in a layer B is given by the same expression with
the subscript A replaced by B.
The boundary condition on the interface 
at the position $x=l$ is given by 
\begin{eqnarray}
&& E_{\rm A}^{(1)}e^{i k_{\rm A} l}
 + E_{\rm A}^{(2)}e^{-i k_{\rm A} l}
= E_{\rm B}^{(1)} e^{i k_{\rm B} l} 
+ E_{\rm B}^{(2)} e^{- i k_{\rm B} l},
\nonumber\\
&&n_{\rm A} (E_{\rm A}^{(1)} e^{i k_{\rm A} l}
-E_{\rm A}^{(2)} e^{-i k_{\rm A} l})
=E_{\rm B}^{(1)} e^{i k_{\rm B} l}
-E_{\rm B}^{(2)} e^{-i k_{\rm B} l}.
\label{BC}
\end{eqnarray}
By introducing the following variables
\begin{eqnarray}
E_+=E^{(1)}+E^{(2)},   
\quad
E_-=(E^{(1)}-E^{(2)})/i,
\label{eq:epm}
\end{eqnarray}
(\ref{BC}) is recast into
\begin{eqnarray}
\left(
\begin{array} {c} 
 E_+ \\ 
 E_-
\end{array}
\right)_{\rm A}
  = {\cal T}^{-1}(n_{\rm A} k l) {\cal T}_{\rm AB}
{\cal T}(k l)
  \left(
\begin{array} {c} 
E_+\\
E_-
\end{array}
 \right)_{\rm B}.
\label{ET}
\end{eqnarray}
Here ${\cal T}(\delta)$ and ${\cal T}_{\rm AB}$ are transfer matrices given by
\begin{eqnarray}
{\cal T}(\delta)=
\left(
\begin{array} {cc} 
 \cos\delta & -\sin\delta \\ 
\sin\delta & \cos\delta 
\end{array}
 \right),
\quad
{\cal T}_{\rm AB}=
\left(
\begin{array} {cc}
1 & 0  \\
 0 & 1/n_{\rm A}
\end{array} 
\right), 
\label{transfer}
\end{eqnarray}
which represent light propagation within a layer and across an interface 
$A \leftarrow B$, respectively.
In a similar manner, from the boundary condition on the interface 
at $x=l+d$, we have  
\begin{eqnarray}
\left(
\begin{array} {c} 
 E_+ \\ 
 E_-
\end{array}
\right)_{\rm B'}
  = {\cal T}^{-1}( k (l+d)) {\cal T}_{\rm BA}
{\cal T}(n_{\rm A} k (l+d))
  \left(
\begin{array} {c} 
E_+\\
E_-
\end{array}
 \right)_{\rm A},
\label{ET1_2}
\end{eqnarray}
where ${\cal T}_{\rm BA}={\cal T}^{-1}_{\rm AB}$ is the transfer matrix
representing light propagation across an interface $B \leftarrow A$.
Combining (\ref{ET}) with (\ref{ET1_2}), we obtain
\begin{eqnarray}
\left(
\begin{array} {c} 
 E_+ \\ 
 E_-
\end{array}
\right)_{\rm B'}
= {\cal T}^{-1}(k(l+d)){\cal T}_{\rm BA}
{\cal T}(n_{\rm A}kd) {\cal T}_{\rm AB}
{\cal T}(kl)
\left(
\begin{array} {c} 
 E_+ \\ 
 E_-
\end{array}
\right)_{\rm B}.
\label{ET1_3}
\end{eqnarray}
Here note that, for a layer A with thickness $d$,
the phase $\delta$ is given by $\delta=n_{\rm A} k d$,
and for a layer B with thickness $d$,
the phase $\delta$ is given by $\delta= k d$.

We are now ready to consider light propagation through C$_j$ in
Fig.\ref{fig:setup}.
Repeating the similar procedure above, 
we obtain the following formula for light propagation through C$_{j}$.
\begin{eqnarray}
\left(
\begin{array} {c} 
 E_+ \\ 
 E_-
\end{array}
\right)_{\rm R}
  =e^{-ik}{\cal M}_j (k)
  \left(
\begin{array} {c} 
E_+\\
E_- 
\end{array}
 \right)_{\rm L},
\label{ET2}
\end{eqnarray}
where $E_{\pm}$ is defined by (\ref{eq:epm}) for $E_{\rm L}^{(1)}$, $E_{\rm
L}^{(2)}$ and $E_{\rm R}^{(1)}$.
The real matrix ${\cal M}_j(k)$ is obtained recursively,
\begin{equation}
 {\cal M}_{j+1}(k)={\cal M}_j\left(k/3 \right) 
{\cal T}\left(k/3\right) 
{\cal M}_j\left(k/3\right), 
\label{MT}
\end{equation}
with the initial condition 
$ {\cal M}_0(k)={\cal T}_{\rm BA} {\cal T}_{\rm A} (n_{\rm A} k) 
{\cal T}_{\rm AB}$.
Using the initial condition and the recursive relation, we can show
\begin{eqnarray}
\det {\cal M}_j(k) = 1,
\quad
({\cal M}_j(k))_{11}=({\cal M}_j(k))_{22}.
\end{eqnarray}

By eliminating $E_L^{(2)}$ from (\ref{ET2}),
the ratio of the amplitude of the transmitted light $E_R^{(1)}$ to 
that of the incident light $E_L^{(1)}$ for C$_j$ is obtained as 
\begin{eqnarray}
\frac{E_R^{(1)}}{E_L^{(1)}}
=\frac{2 e^{-i k}}{2a_j+i(b_j-c_j)},
\label{E1}
\end{eqnarray}
where $a_j(k)$, $b_j(k)$ and $c_j(k)$ are components of
${\cal M}_j(k)$
\begin{eqnarray}
{\cal M}_j(k) =\left(
\begin{array} {cc}
a_j(k) & b_j(k)  \\
c_j(k) & a_j(k)
\end{array}
\right),
\quad
a_j(k)^2-b_j(k)c_j(k)=1.
\label{eq:mcomponent}
\end{eqnarray}
Therefore the transmission coefficient  $T_j\equiv |E_{\rm R}^{(1)}|^2/|E_{\rm
L}^{(1)}|^2$ of ${\rm C}_j$ is given by  
\begin{equation}
T_j(k)
=\frac{4}{|{\cal M}_j(k)|^2+2}, 
\label{MJ_pre}
\end{equation}
where $|{\cal M}_j(k)|^2\equiv 2a_j(k)^2+b_j(k)^2+c_j(k)^2$.

\begin{figure}
 \begin{center}
  \includegraphics[width=7.0cm,clip]{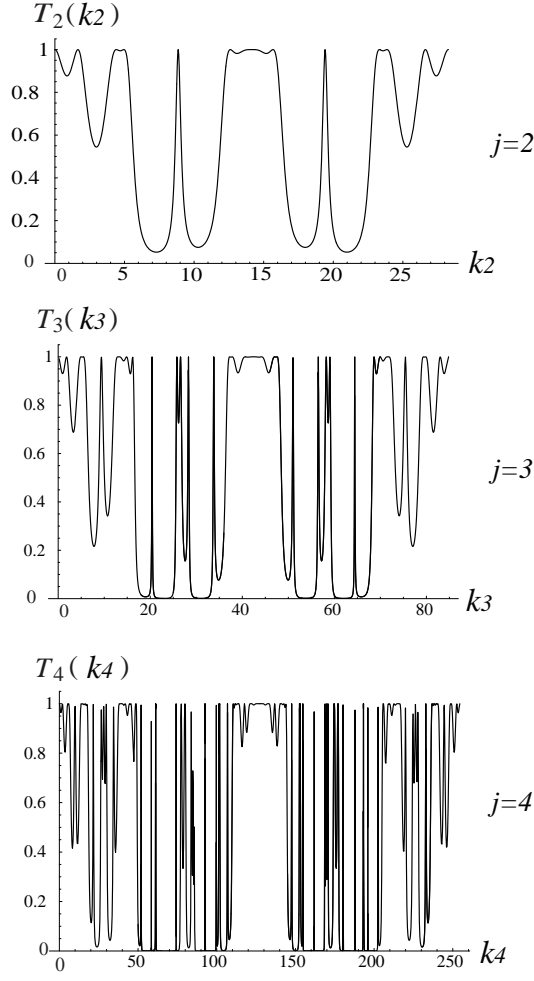}
  \caption{\label{fig:transmission}Transmission coefficients $T_j$ as
  functions of the wave number $k$ for C$_2$, C$_3$, and C$_4$ 
(from top to bottom) with $n_{\rm A}=2.0$.
  The range of the wave number
  $k$ for C$_j$ is $0 \le k \le 3^j \pi$. Note that the
  horizontal axes are rescaled by a factor of three.} 
 \end{center}
\end{figure}

In Fig.\ref{fig:transmission} we illustrate a typical example of
transmission coefficients as functions of the wave number for C$_2$,
C$_3$, and C$_4$. 
As is shown clearly, we have a scaling behavior of the transmission
coefficients: If we multiply the wave number by three as one
generation increases, we have a similar structure in the
transmission coefficients.
To describe the scaling behavior properly,
we introduce the rescaled wave number $k_j=3^{j}\kappa$ for C$_{j}$,  
then from (\ref{MT}), we have
\begin{eqnarray} 
{\cal M}_{j+1}(k_{j+1})={\cal M}_{j}(k_j)
{\cal T}(3^{j}\kappa){\cal M}_{j}(k_j),
\quad
{\rm det}{\cal M}_j(k_j)=1,
\quad
({\cal M}_j(k_j))_{11}=({\cal M}_j(k_j))_{22},
\label{MT3}
\end{eqnarray}
which can be regarded as the {\it renormalization-group equation} describing
the scaling behavior.
The ``renormalized'' transmission coefficient $T_j(k_j)$ for C$_j$ is
given by
\begin{eqnarray}
T_j(k_j)=\frac{4}{|{\cal M}_j(k_j)|^2+2}. 
\label{MJ}
\end{eqnarray}

\section{Chaotic propagation, complete transmission, 
and complete reflection of light}
\label{sec:results}

In this section we study the scaling behaviors of the
wave propagation by using the renormalization-group equation (\ref{MT3}).  
Using analytical and numerical methods,  
we find two different scaling behaviors depending on $\kappa$
of the rescaled wave number $k_j=3^j\kappa$:
a) For $\kappa=(m/2\cdot 3^q)\pi$ with integers $m$ and $q$, the
renormalization-group equation reduces to the logistic map describing
a chaotic behavior. The renormalized transmission coefficient 
$T_j(k_j)$ is very
sensitive to parameters of the system, and it changes drastically as the
generation increases.
b) For the other $\kappa$,  it  will be found numerically that the renormalized
transmission  coefficient $T_j(k_j)$ eventually flows into either $T=1$
or $T=0$ as $j\rightarrow \infty$.

\subsection{$\kappa=(m/2\cdot3^{q})\pi$}
\label{sec:chaos}

\begin{figure}
\begin{center}
\includegraphics[width=8.5cm,clip]{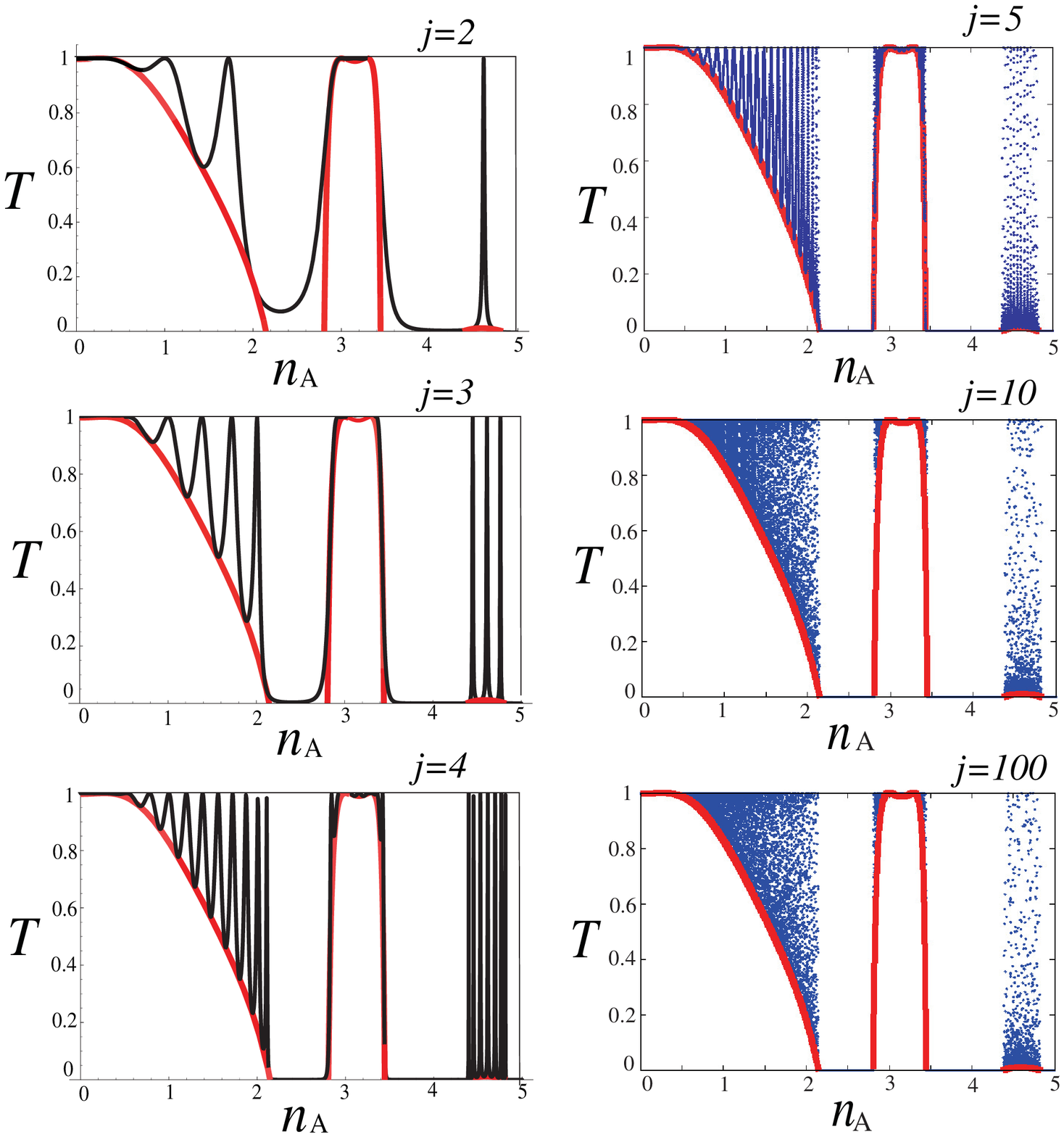}
   \caption{\label{fig:chaos}
(Color online) Transmission coefficients $T_j$ as functions of $n_{\rm A}$
 for $\kappa=\pi/3$. 
The generations of the left panel are $j=2$, $j=3$, and $j=4$.
We also show transmission coefficients for $j=5$, $j=10$, and $j=100$
in the right panel.
We plot $T_{\rm min}$ as a function of $n_{\rm A}$ by the thick red line.
} 
\label{fig:0.33PI} 
\end{center}
\end{figure}
\begin{figure}
  \begin{center}
  \includegraphics[width=6.5cm]{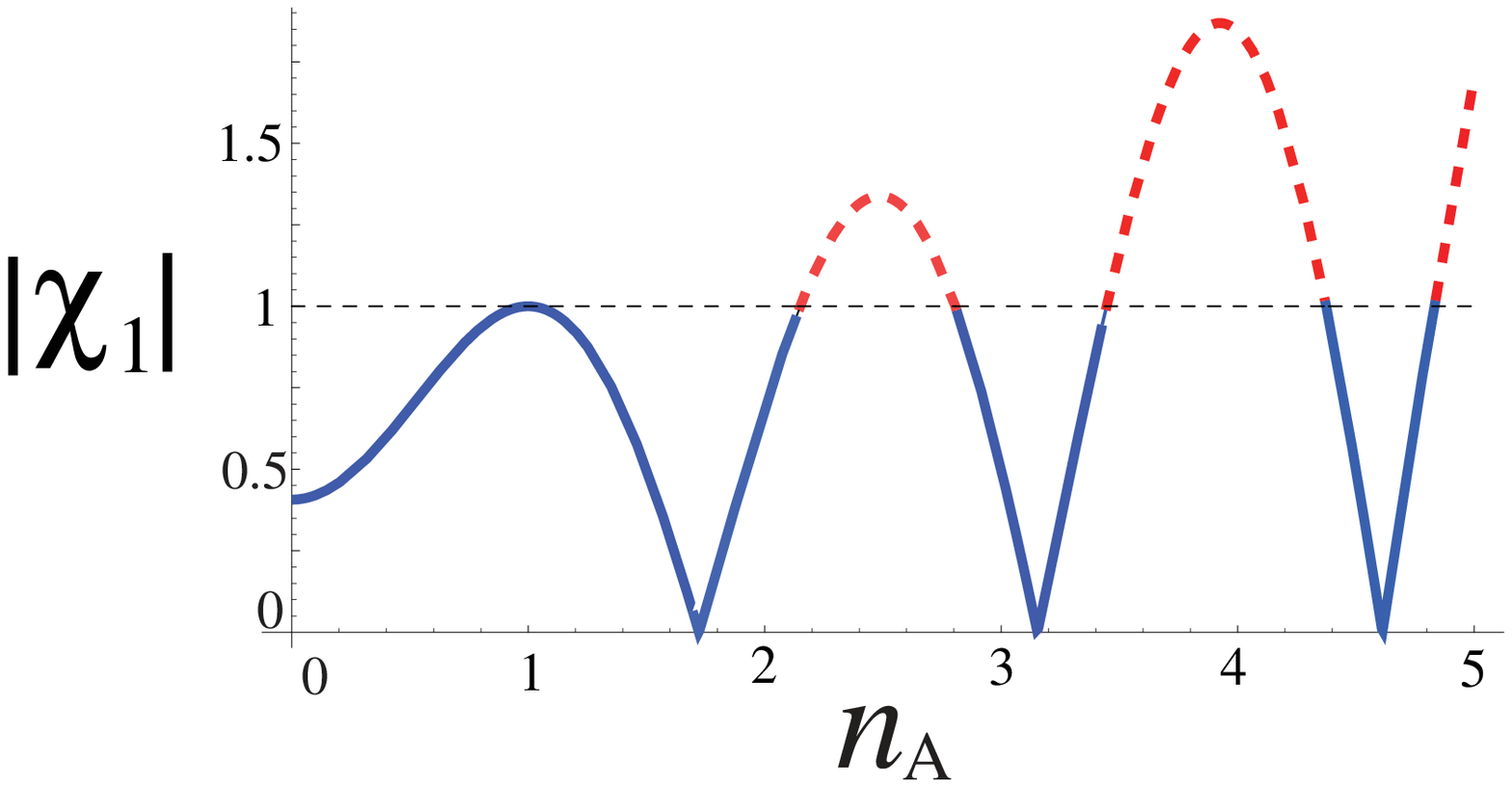}
  \caption{
(Color online) The value of $|\chi_1|$ as a function of $n_{\rm A}$ 
for $\kappa=\pi/3$.
For $n_{\rm A}$ which gives $|\chi_1|<1$ (solid lines), 
the transmission coefficients show chaotic behaviors.
For $n_{\rm A}$ which gives $|\chi_1|>1$ (dashed lines), 
the transmission coefficients flow into $T=0$ for $j\to\infty$.}
  \label{fig:0.33PI_x1}
 \end{center}
\end{figure}

\begin{figure}
 \begin{center}
\includegraphics[width=8.5cm,clip]{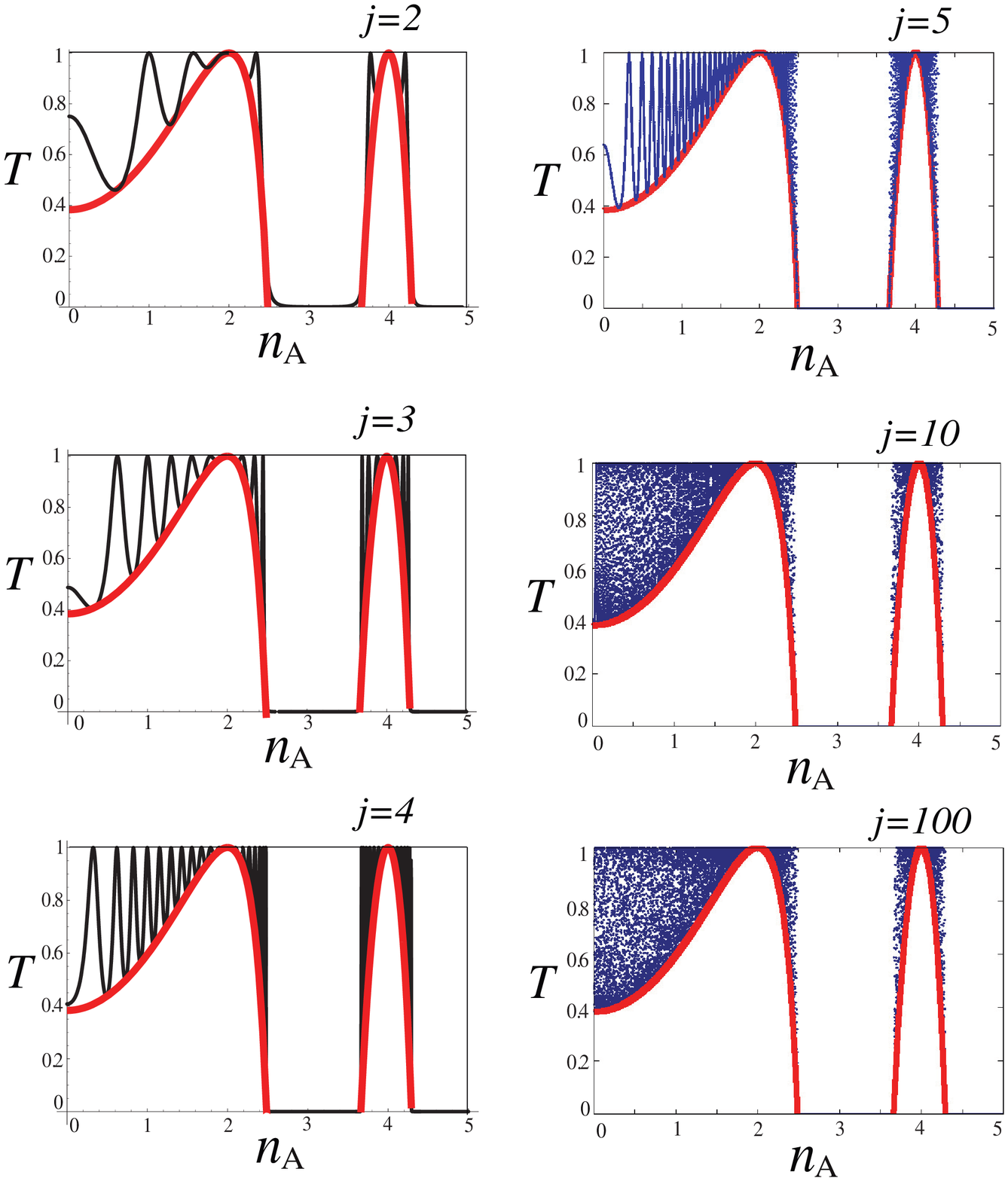}
   \caption{(Color online) Transmission coefficients $T_j$ as
   functions of $n_{\rm A}$ for $\kappa=\pi/2$. 
The generations of the left panel are $j=2$, $j=3$, and $j=4$.
We also show transmission coefficients for $j=5$, $j=10$, and $j=100$
in the right panel.
We plot $T_{\rm min}$ as a function of $n_{\rm A}$ by the thick red line.
} 
\label{fig:0.5PI}
 \end{center}
\end{figure}
\begin{figure}
  \begin{center}
  \includegraphics[width=6.5cm]{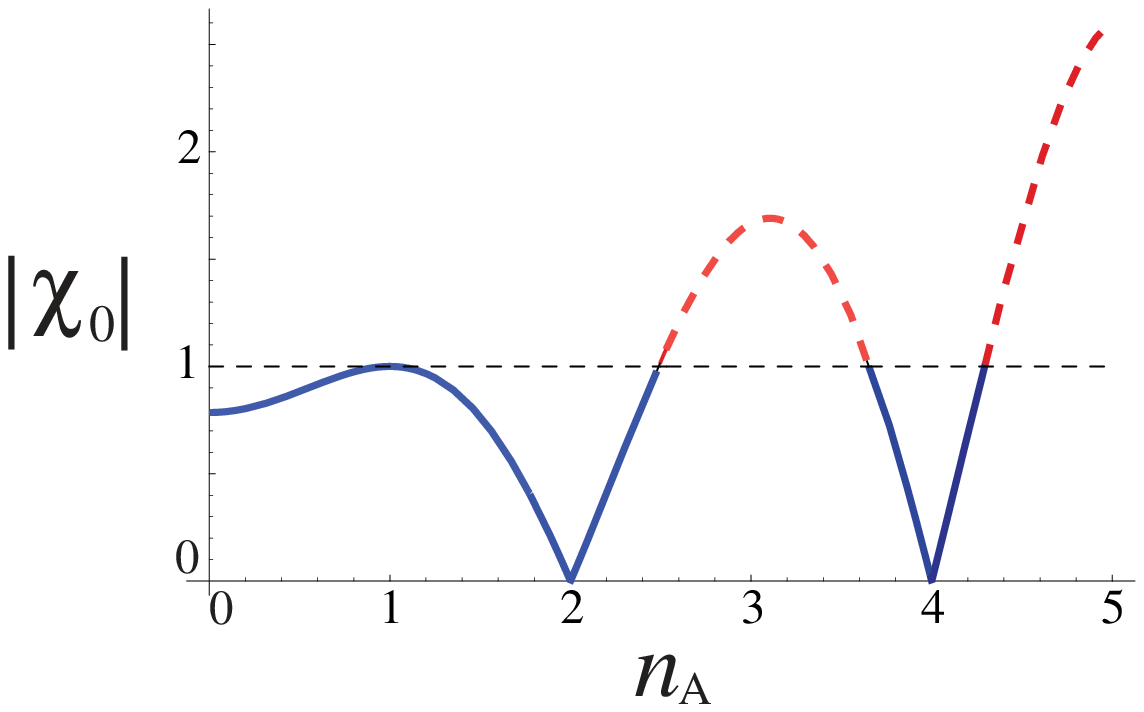}
 \end{center}
  \caption{(Color online) The value of $|\chi_0|$ as a function of $n_{\rm A}$ 
for $\kappa=\pi/2$.
For $n_{\rm A}$ which gives $|\chi_0|<1$ (solid lines), 
the transmission coefficients show chaotic behaviors.
For $n_{\rm A}$ which gives $|\chi_0|>1$ (dashed lines), 
the transmission coefficients flow into $T=0$ for $j\to\infty$.}
  \label{fig:0.5PI_y0}
 \end{figure}

Let us first suppose that $m$ is even, $m=2n$.
In this case, ${\cal T}(3^j\kappa)$ in (\ref{MT3}) takes a unique
value ${\cal T}(n\pi)$ for $j\ge q$
\footnote{
Layers B in C$_j$ have $j$ different lengths, $1/3^j,
1/3^{j-1}, \cdots 1/3$, thus in general they give $j$ different transfer
matrices, ${\cal T}(\kappa), {\cal T}(3\kappa), \cdots {\cal T}(3^{j-1}
\kappa)$.
But for  $\kappa=(n/3^q)\pi$ with $j\ge q$,  
 ${\cal T}(\delta)$'s in B's larger than $1/3^{j-q}$
are equal to ${\cal T}(n \pi)$.},
which enables us to solve (\ref{MT3}) analytically.
By defining ${\cal N}_j ={\cal T}(3^{j+1}\kappa){\cal M}_j$, (\ref{MT3}) is
recast into
\begin{eqnarray}
{\cal N}_{j+1}={\cal N}_j^2,~(j\ge q),
\quad
{\rm det} {\cal N}_j=1,
\quad
{\cal N}_j={\cal T}(n\pi){\cal M}_j,
\label{NT}
\end{eqnarray}
and from the relation $|{\cal N}_j|^2=|{\cal M}_j|^2$, the renormalized
transmission coefficient $T_j(k_j)$ is rewritten as 
\begin{eqnarray}
T_j=\frac{4}{|{\cal N}_j|^2+2}. 
\label{MJ2}
\end{eqnarray}

To solve (\ref{NT}), we rewrite ${\cal N}_j$ in terms of the Pauli
matrices $\sigma_i$ $(i=1,2,3)$, 
\begin{eqnarray}
{\cal N}_j
=&\chi_j{\bm 1}+{\bm \alpha}_j\cdot {\bm \sigma}, 
\quad {\bm \alpha}_j=(\alpha_j^{(1)},i\alpha_j^{(2)},\alpha_j^{(3)}),
\label{eq:chialpha}
\end{eqnarray}
with real $\chi_j$ and $\alpha_j^{(i)}$ $(i=1,2,3)$. 
From (\ref{NT}), we have 
\begin{eqnarray}
\chi_{j+1}=\chi_j^2+{\bm \alpha}_j^2,
\quad
{\bm \alpha}_{j+1}=2 \chi_j {\bm \alpha}_{j},
\quad
\chi_j^2-{\bm \alpha}_j^2=1,
\quad
(j\ge q),
\label{detn2}
\end{eqnarray}
where ${\bm \alpha}_j^2=(\alpha_j^{(1)})^2-(\alpha_j^{(2)})^2
+(\alpha_j^{(3)})^2$.
By eliminating ${\bm \alpha}^2_j$ in (\ref{detn2}), 
the map for $\chi_j$ is obtained,
\begin{equation}
\chi_{j+1}=2\chi_j^2-1,
\quad
(j\ge q),  
\label{xjlogistic}
\end{equation}
and in terms of the solution $\chi_j$ of (\ref{xjlogistic}),  ${\bm \alpha}_j$
and $T_j$ are represented as
\begin{eqnarray}
{\bm \alpha}_j=\prod_{k=q}^{j-1}(2\chi_k){\bm \alpha}_q, 
\quad
T_j
=\frac{1-(\lambda^2+\eta^2)}{1- (\lambda^2+\eta^2)\chi_j^2},
\quad (j\ge q).
\label{aequation}
\end{eqnarray}
Here $\lambda$ and $\eta$ are the constants of motion in (\ref{detn2}), 
\begin{eqnarray}
\frac{\alpha_j^{(3)}}{\alpha_j^{(2)}}=\frac{\alpha_q^{(3)}}{\alpha_q^{(2)}}
\equiv\lambda,
\quad 
\frac{\alpha_j^{(1)}}{\alpha_j^{(2)}}=\frac{\alpha_q^{(1)}}{\alpha_q^{(2)}}
\equiv\eta,
\quad (j\ge q).
\label{constants_1}
\end{eqnarray}
For $|\chi_q|\le 1$, they satisfy $\lambda^2+\eta^2\le 1$, and for
$|\chi_q|> 1$, $\lambda^2+\eta^2 >1$.

Equations (\ref{xjlogistic}) and (\ref{aequation}) can be solved analytically.
For $0\le |\chi_q|< 1$, the solution is 
\begin{eqnarray}
&&\chi_j=\cos[2^{j-q}\cos^{-1}{\chi_q}],
\quad
{\bm \alpha}_j=\frac{\sin[2^{j-q} \cos^{-1} \chi_q]}{\sin[\cos^{-1} \chi_q]}
  {\bm \alpha}_q,
\nonumber\\
&&T_j=\frac{1-(\lambda^2+\eta^2)}{1-(\lambda^2+\eta^2)
 \cos^2[2^{j-q}\cos^{-1}\chi_q]}, 
\label{eq:solution1}
\end{eqnarray}
and for $|\chi_q|\ge 1$, 
\begin{eqnarray}
&&\chi_j=\cosh[2^{j-q}\cosh^{-1} |\chi_q|],
\quad
{\bm \alpha}_j=\frac{\sinh[2^{j-q} \cosh^{-1} |\chi_q|]}
 {\sinh[ \cosh^{-1} |\chi_q|]} {\bm \alpha}_q,
\nonumber\\
&&T_j
=\frac{(\lambda^2+\eta^2)-1}
{(\lambda^2+\eta^2)\cosh^2[2^{j-q}\cosh^{-1}|\chi_q|]-1}.
\label{eq:solution2}
\label{eq:solution2}
\end{eqnarray}

For $|\chi_q|< 1$, the transmission coefficient $T_j$ shows sensitive
dependence on parameters of the system. 
To illustrate this, we show the transmission coefficients for $q=1$
and $n=1$ case, (namely $\kappa=\pi/3$), as functions of $n_{\rm A}$ in
Fig.\ref{fig:0.33PI}. In this case,
$\chi_{q=1}$ is given by
\begin{eqnarray}
\chi_1=-\frac{1}{2}\cos \left(\frac{2n_{\rm A}\pi}{3}\right)
+\frac{\sqrt{3}}{4}\left(n_{\rm A}+\frac{1}{n_{\rm A}} \right)
\sin\left(\frac{2n_{\rm A}\pi}{3}\right),  
\label{chi_1}
\end{eqnarray}
and the region of $n_{\rm A}$ with $|\chi_1|< 1$ is shown in
Fig.\ref{fig:0.33PI_x1}.
It is found clearly that the transmission coefficients are very sensitive to $n_{\rm A}$ in the
region where $|\chi_1|< 1$.

This sensitivity to parameters of the system can be properly
understood by
introducing a new variable $X_j=1-\chi_j^2$.
In terms of the new variable, (\ref{xjlogistic}) reduces to
the logistic map with $r=4$,
\begin{eqnarray}
X_{j+1}=r X_j(1-X_j),    
\end{eqnarray}
which has been studied extensively in the context of chaos \cite{LY75,
May76, Fei79,Fei79_2}.
The logistic map with $r=4$ in the interval $0<X<1$ is known to be very
sensitive to the initial condition of the system, which implies that $T_j$ for
$|\chi_q|<1$ also has the same chaotic property.

A similar chaotic behavior of the transmission coefficient
also appears for odd $m=2n+1$,
\begin{eqnarray}
\kappa=\frac{2n+1}{2\cdot 3^{q}}\pi.
\label{coherent2}
\end{eqnarray}
The matrix ${\cal T}(3^j\kappa)$ in (\ref{MT3})
is now $(-1)^{j-q}{\cal T}((n+1/2)\pi)$  for $j\ge q$, so
${\cal N}_j(\equiv {\cal T}(3^{j+1}\kappa){\cal M}_j)$ becomes ${\cal N}_j=(-1)^{j+1-q}{\cal
T}((n+1/2)\pi){\cal M}_j$.
For this ${\cal N}_j$, we have the same equation as (\ref{NT}), 
\begin{eqnarray}
{\cal N}_{j+1}={\cal N}_j^2,
\quad
{\rm det}{\cal N}_j=1,
\end{eqnarray}
and the same expression of the transmission coefficient as (\ref{MJ2}),
\begin{eqnarray}
T_j=\frac{4}{|{\cal N}_j|^2+2}.  
\end{eqnarray}
By using $\chi_j$, ${\bm \alpha}_j$, $\lambda$ and $\eta$ defined in
(\ref{eq:chialpha}) and (\ref{constants_1}), 
the same solutions (\ref{eq:solution1}) and (\ref{eq:solution2}) are obtained.
Thus we have the same class of chaotic behavior.
As an example, we show the transmission coefficients $T_j$ for
$\kappa=\pi/2$ ($n=q=0$ in (\ref{coherent2})) as functions of $n_{\rm
A}$ in Fig.\ref{fig:0.5PI}.
For $\kappa=\pi/2$, $\chi_{q=0}$ is given by
\begin{eqnarray}
\chi_0=\frac{1}{2}\left(n_{\rm A}
+\frac{1}{n_{\rm A}}\right)\sin\left(\frac{n_{\rm A}\pi}{2}\right).
\label{x_initial}
\end{eqnarray}
and the region with $|\chi_0|< 1$ is shown in Fig.\ref{fig:0.5PI_y0}. 
We find again that $T_j$ is very sensitive to $n_{\rm A}$ in the region
with $|\chi_0|< 1$.

In the chaotic region, $|\chi_q|<1$, the transmission
coefficient $T_j$ has the lower bound $T_{\rm min}$
\begin{eqnarray}
T_{\rm min}=1-(\lambda^2+\eta^2).
\end{eqnarray}
For $\kappa=\pi/3$,  $\lambda$ and $\eta$ are given by
\begin{eqnarray}
\lambda=0,
\quad
\eta=\frac{-\sqrt{3}(n_{\rm A}^2-n_{\rm A}^{-2})
\sin^2(\pi n_{\rm A}/3)+(n_{\rm A}-n_{\rm A}^{-1})
\sin(2\pi n_{\rm A}/3)}
{\sqrt{3}(n_{\rm A}+n_{\rm A}^{-1})^2\sin^2(\pi n_{\rm A}/3)-(n_{\rm
A}+n_{\rm A}^{-1})\sin(2\pi n_{\rm A}/3)-2\sqrt{3}},
\end{eqnarray}
and for $\kappa=\pi/2$,
\begin{eqnarray}
\lambda=\frac{1}{2}\left(n_{\rm A}-\frac{1}{n_{\rm A}}\right)
\tan\left(\frac{\pi n_{\rm A}}{2}\right),
\quad
\eta=0.
\label{constants}
\end{eqnarray}
The resultant $T_{\rm min}$'s as functions of $n_{\rm A}$ are also depicted
in Figs.{\ref{fig:0.33PI}} and \ref{fig:0.5PI}, respectively.

From the analytic solution (\ref{eq:solution2}), we notice that
outside the chaotic region, $|\chi_q|\ge 1$, $T_j$ decreases monotonically
as the generation $j$ increases.
Thus we have complete reflection $T_j=0$ for $j\to\infty$.

\subsection{Other $\kappa$}
\begin{figure}
 \begin{center}
   \includegraphics[width=6.5cm,clip]{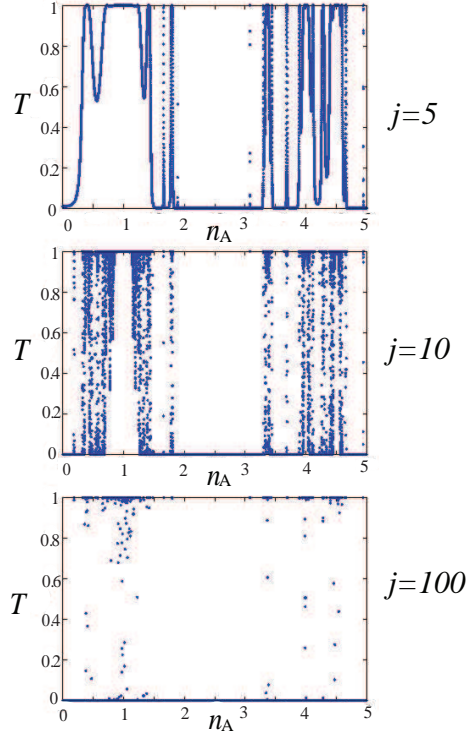}
   \caption{\label{fig:nonchaos}
(Color online) Transmission coefficients $T_j$ as
   functions of $n_{\rm A}$ for $\kappa=\pi/4$.
 The generations are $j=5$, $j=10$, and $j=100$ (from top to bottom).} 
 \end{center}
\end{figure}

For $\kappa\neq (m/2\cdot 3^q)\pi$, we numerically find that the
transmission coefficient $T_j$ flows into either $T=1$ or $T=0$ as the
generation $j$ increases.
A typical example of our numerical results is shown 
in Fig.\ref{fig:nonchaos}.
It is also found that if complete transmission of light occurs 
at a certain generation, light always transmits completely in the
following generations.
To see this, let us use the component representation of 
${\cal M}_j(k_j)$ as (\ref{eq:mcomponent}),
\begin{eqnarray}
{\cal M}_j(k_j)=
\left(
\begin{array}{cc}
a_j & b_j \\
c_j & a_j
\end{array}
\right), 
\end{eqnarray}
which satisfies $a_j^2-b_jc_j=1$.
If $T_j=1$, $2a_j^2+b_j^2+c_j^2=2$ from (\ref{MJ}), thus ${\cal M}_j$ can be parameterized as
\begin{eqnarray}
{\cal M}_j(k_j)=
\left(
\begin{array}{cc}
\cos \delta_j & -\sin \delta_j\\
\sin\delta_j & \cos \delta_j
\end{array}
\right),
\quad a_j=\cos \delta_j,
\quad b_j=-c_j=-\sin\delta_j.
\end{eqnarray}
Substituting this into (\ref{MT3}) yields
\begin{eqnarray}
{\cal M}_{j+1}(k_{j+1})=
\left(
\begin{array}{cc}
\cos (2\delta_j+k_j) & -\sin(2\delta_j+k_j)\\
\sin(2\delta_j+k_j)  & \cos(2\delta_j+k_j)
\end{array}
\right), 
\end{eqnarray}
which implies $T_{j+1}=1$.

\begin{figure}
 \begin{center}
  \includegraphics[width=6.5cm,clip]{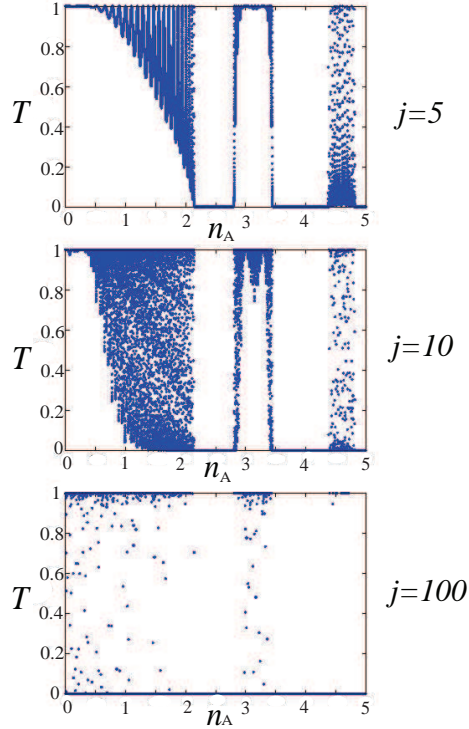}
  \caption{\label{fig:nearchaos}
(Color online) Transmission coefficients $T_j$ 
as functions of $n_{\rm A}$ for $\kappa=\pi/3+10^{-4}$. 
 The generations are $j=5$, $j=10$, and $j=100$ (from top to bottom).} 
 \end{center}
\end{figure}
When $\kappa$ is near $\kappa=(m/2\cdot 3^q)\pi$, the
transmission coefficient is found to show transition from the
chaotic behavior presented in the previous subsection:
The chaotic behavior appears in first few generations, 
but finally the transmission coefficient flows into 
either complete transmission or complete reflection. 
For example, the transmission coefficient $T_j$ for 
$\kappa=\pi/3+10^{-4}$
is shown in Fig.\ref{fig:nearchaos}.
In the fifth generation ($j=5$), the transmission coefficient shows
a chaotic behavior indistinguishable from that for 
$\kappa=\pi/3$ (Fig.\ref{fig:0.33PI}),
however, in the tenth generation ($j=10$),
it shows a different behavior 
 from that for $\kappa=\pi/3$,
 then finally in the hundredth generation ($j=100$),
 it flows into either $T=1$ or $T=0$.

\subsection{Optical experiments}

The chaotic behaviors of the transmission coefficients
presented in Sec.\ref{sec:chaos} can be observed experimentally. 
As was shown in Figs.\ref{fig:chaos} and \ref{fig:0.5PI},
the chaotic characteristics in the propagation are already evident in
the first few generations of the Cantor multilayers,
where the transmission coefficients oscillate rapidly 
as functions of $n_{\rm A}$.
(This oscillation reflects the stretching and the folding process 
of the logistic map.)
Therefore, the finite generations of the Cantor multilayers are
sufficient to observe the chaotic behaviors. 
Experimentally, the finite generations of C$_j$ can be fabricated by
using the vacuum deposition on a glass substrate \cite{wg94,th94,av02}.
In a similar manner as the Fibonacci multilayers \cite{wg94,th94}, the
Cantor multilayers are prepared from SiO${}_2$ and TiO${}_2$ films.
Na$_3$AlF$_6$ and ZnS are also utilized for the fabrication
of C$_j$ \cite{av02}.
Although the wave number $k$ needs to satisfy the condition
$k=3^j\kappa$ with $\kappa=(m/2\cdot 3^q)\pi$ ($m$ and $q$ are integers)
in order to observe the chaotic behaviors on the 
$j$th generation of the Cantor
multilayers, this condition can be met by using tunable lasers.

From the argument in Sec.\ref{sec:chaos}, the indices of the refraction of
A and B should be set in the region with $0\le |\chi_q| < 1$.
Suppose that the index of refraction of B is 1,
then $\chi_q$ is given by (see Eq.(\ref{eq:chialpha}))
\begin{eqnarray}
\chi_q
=\frac{1}{2}{\rm Tr}({\cal T}(3^{q+1}\kappa){\cal M}_q),
\end{eqnarray}
where ${\cal M}_q(\equiv {\cal M}_q(k_q))$ is the solution of Eq.(\ref{MT3}).
For example, in the case of $\kappa=\pi/3$ ($\kappa=\pi/2$),
$\chi_{q=1}$ ($\chi_{q=0}$) as a function of the index of the refraction of A,
$n_{\rm A}$, is given by Eq.(\ref{chi_1}) (Eq.(\ref{x_initial})).
As was illustrated in Fig.\ref{fig:0.33PI_x1} (Fig.\ref{fig:0.5PI_y0}),
the condition $0 \le |\chi_q|< 1$ can be met in a broad region of
$n_{\rm A}$ without fine tuning. 
Therefore, the detection of the chaotic behaviors is feasible for
the current optical experiments.



The most impressive chaotic behavior is obtained if $n_{\rm A}$ satisfies 
$|\chi_q|\sim 1$.
Near $n_{\rm A}$ satisfying $|\chi_q|=1$,
all values of $T$ appear in a very narrow region of $n_{\rm A}$.
In particular, complete transmission and (almost) complete reflection 
are nearby in the narrow region.
By controlling $n_{\rm A}$, 
the Cantor multilayers with $|\chi_q|\sim 1$ could be used 
as fast switching devices.




\section{Local scaling behavior of transmission coefficients
near complete transmission}
\label{sec:results2}

In the previous section, we found two classes of behaviors of
the transmission coefficients where they remain finite:
chaotic behaviors and complete transmission.
In this section, we focus on a behavior of the transmission coefficient
near complete transmission.
As we showed in the previous section,
if we have complete transmission of light at a certain generation,
light always transmits completely after the generation.
Thus we can consider that complete transmission is 
a fixed point of the renormalization-group equation (\ref{MT3}).
According to the renormalization-group theory,
the existence of a fixed point implies that scalings are found
around it \cite{Kadanoff}, which turns out to be also true in our problem.
In the following, on the basis of the renormalization-group equation,
it will be shown that if complete transmission is achieved at a wave
number $\kappa^*$ with a rational $\kappa^*/\pi$, then the transmission
coefficient $T_j$ around the wave number exhibits a local
scaling behavior which is distinct from the 
global scaling illustrated in Fig.\ref{fig:transmission}.
Moreover, we will present the analytic expression of the scaling function.

\begin{figure}
  \begin{center}
   \includegraphics[width=5.0cm,clip]{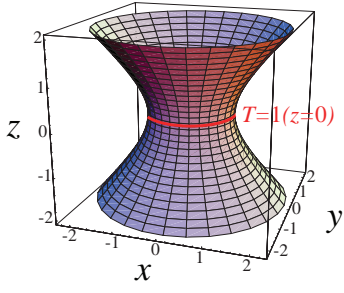}
  \end{center}
  \caption{
(Color online) Orbits of (\ref{MT3}) are constrained 
on the manifold $x^2+y^2=z^2+1$.
   Complete transmission $T=1$ corresponds to $z=0$.}
  \label{fig:hyperboloid}
\end{figure}

In order to analyze the scaling behavior of $T_j$ near $T=1$,
it is convenient to introduce new variables $x_j\equiv a_j(k_j)$,
$y_j\equiv (b_j(k_j)-c_j(k_j))/2$ and $z_j\equiv (b_j(k_j)+c_j(k_j))/2$
for the matrix elements of ${\cal M}_j(k_j)$,
\begin{eqnarray}
{\cal M}_j(k_j) =\left(
\begin{array} {cc}
a_j(k_j) & b_j(k_j)  \\
c_j(k_j) & a_j(k_j)
\end{array}
\right),
\quad k_j=3^j\kappa.
\end{eqnarray}
Since $a_j(k_j)$, $b_j(k_j)$ and $c_j(k_j)$ satisfy
$a_j(k_j)^2-b_j(k_j)c_j(k_j)=1$,
the new variables $(x_j,y_j,z_j)$ are constrained 
on the manifold $x_j^2+y_j^2=1+z_j^2$
shown in Fig.\ref{fig:hyperboloid}.
By rewriting  $x_j$ and $y_j$ as
\begin{eqnarray}
x_j=\sqrt{z_j^2+1}\cos \varphi_j,
\quad
y_j=\sqrt{z_j^2+1}\sin \varphi_j,
\label{xy_varphi}
\end{eqnarray}
the map (\ref{MT3}) is recast into
\begin{eqnarray}
z_{j+1}&=&2z_j\sqrt{z_j^2+1}\cos(\varphi_j-3^j \kappa),
\nonumber\\ 
\sqrt{z_{j+1}^2+1}e^{i\varphi_{j+1}}&=&(z_j^2+1)e^{i(2 \varphi_j-3^j \kappa)}
                                        +z_j^2 e^{i3^j \kappa},
\label{mainmap2_z}
\end{eqnarray}
and the transmission coefficient $T_j$ in (\ref{MJ}) is rewritten as 
\begin{eqnarray}
T_j=\frac{1}{x_j^2+y_j^2}=\frac{1}{1+z_j^2}.
\label{transmission_zj}
\end{eqnarray}
The complete transmission $T_j=1$ is achieved when $z_j=0$.
From (\ref{mainmap2_z}), it can be found that this occurs either
if 1) $z_l=0$ with $z_{l-1}\neq 0$ for an integer $l$ ($1\le l \le j$), or 2) $z_0=0$.

First consider the case 1) in detail.
Suppose that $z_l=0$ with $z_{l-1}\neq 0$ is realized for $\kappa=\kappa_l^*$.
Then from the first equation of (\ref{mainmap2_z}), it is found that the case
1) is possible only if $\cos(\varphi_{l-1}-3^{l-1}\kappa_l^*)=0$.
Thus $\varphi_{l-1}$ is given by
\begin{eqnarray}
\varphi_{l-1}=3^{l-1}\kappa_l^*+\frac{2n+1}{2}\pi,
\end{eqnarray}
with an integer $n$.
Then using (\ref{mainmap2_z}) with $j=l-1$ and $z_{l}=0$, we have 
\begin{eqnarray}
e^{i\varphi_{l}}&=&
(z_{l-1}^2+1)e^{i(2\varphi_{l-1}-3^{l-1}\kappa_l^*)}
+z_{l-1}^2e^{i3^{l-1}\kappa_l^*}
\nonumber\\
&=&(z_{l-1}^2+1)e^{i(3^{l-1}\kappa_l^*+(2n+1)\pi)}
 +z_{l-1}^2e^{i3^{l-1}\kappa_l^*}
\nonumber\\
&=&-e^{i3^{l-1}\kappa_l^*}. 
\end{eqnarray}
Hence $\varphi_{l}$ is obtained as
\begin{eqnarray}
\varphi_{l}=3^{l-1}\kappa_l^*+\pi,
\quad
(\mbox{mod $2\pi$}).
\label{eq:varphi1}
\end{eqnarray}
To determine $\varphi_j$ for $j\ge l$, we use the second
equation of (\ref{mainmap2_z}).
Since $z_j=0$ for $j\ge l$, the second equation of
(\ref{mainmap2_z}) becomes
\begin{eqnarray}
e^{i\varphi_{j+1}}=e^{i(2\varphi_j-3^j\kappa_l^*)}
\quad (j\ge l),
\label{eq:varphi2}
\end{eqnarray}
which determines $\varphi_j$ for $j\ge l$ completely as,
\begin{eqnarray}
\varphi_j=-(3^{j}-2^{j-l}3^{l})\kappa_l^*+2^{j-l}\varphi_{l} 
\quad
(j\ge l; \mbox{mod $2\pi$}),
\end{eqnarray}
where $\varphi_l$ is given by (\ref{eq:varphi1}).

In a similar manner, we can also solve $\varphi_j$ in the case 2).
Suppose that $z_0=0$ for $\kappa=\kappa_0^*$.
From the following explicit form of ${\cal M}_0$,
\begin{eqnarray}
{\cal M}_0(k_0)={\cal T}_{\rm BA}
 {\cal T}_{\rm A}(n_{\rm A}k_0){\cal T}_{\rm AB}
=\left(
\begin{array}{cc}
\cos n_{\rm A}\kappa & -(1/n_{\rm A})\sin n_{\rm A}\kappa \\
n_{\rm A}\sin n_{\rm A}\kappa & \cos n_{\rm A}\kappa
\end{array}
 \right),
\label{eq:z0}
\end{eqnarray}
we obtain that $z_0=(n_{\rm A}-1/n_{\rm A})\sin(n_{\rm
A}\kappa)/2$.
Therefore, $\kappa_0^*$ is given by $\kappa_0^*=m\pi/n_{\rm A}$ with an
integer $m$.
(Note that the indices of the refraction of A and B are different
from each other, {\it i.e.} $n_{\rm A}\neq 1$.)
Substituting this into (\ref{eq:z0}), we find that $x_0=\cos m\pi$ and
$y_0=0$.
Thus $\varphi_0$ is given by
\begin{eqnarray}
\varphi_0=m\pi,\quad (\mbox{mod $2\pi$}). 
\label{eq:varphi0}
\end{eqnarray}
Since $z_j=0$ for $j\ge 0$,  we have
\begin{eqnarray}
e^{i\varphi_{j+1}}=e^{i(2\varphi_j-3^j\kappa_0^*)}
\quad
(j\ge 0),
\end{eqnarray}
from Eq.(\ref{mainmap2_z}), 
which determines $\varphi_j$ as
\begin{eqnarray}
\varphi_j=-(3^j-2^j)\kappa_0^*+2^j \varphi_0,
\quad
(\mbox{mod $2\pi$}),
\end{eqnarray}
with $\varphi_0=m\pi$.

Now we study behavior of the transmission coefficient $T_j$ near
$T=1$.
For $\kappa$ near $\kappa^*_l$, $z_l$ becomes nonzero but remains small,
so we can neglect higher
order terms of $z_i^2$ $(i\ge l)$ in the map (\ref{mainmap2_z}).
Up to the next leading order, the map (\ref{mainmap2_z}) is approximated by
\begin{eqnarray}
z_{j+1}=2 z_j \cos(\varphi_j-3^j\kappa),
\quad
e^{i\varphi_{j+1}}=e^{i(2\varphi_j-3^j\kappa)}.
\label{z_solution_pre}
\end{eqnarray}
The latter equation in
(\ref{z_solution_pre}) can be solved easily,
\begin{eqnarray}
\varphi_j=-(3^j-2^{j-l}3^l)\kappa+2^{j-l}\varphi_{l},  
\quad
(\mbox{mod $2\pi$}),
\end{eqnarray}
where $\varphi_l$ is given by
\begin{eqnarray}
\varphi_l=
\left\{
\begin{array}{ll}
3^{l-1}\kappa_l^*+\pi+O(z_l), &\mbox{for $l\neq 0$} \\
m\pi+O(z_0), & \mbox{for $l=0$}
\end{array}
\right. .
\quad (\mbox{mod $2\pi$}).
\end{eqnarray}
Here we have determined $\varphi_l$ in a similar manner as
(\ref{eq:varphi1}) and (\ref{eq:varphi0}), but $O(z_l)$ corrections
appear since $z_l\neq 0$ near $\kappa_l^*$. 
Substituting this into the first equation in
(\ref{z_solution_pre}), we obtain 
\begin{eqnarray}
z_{j+1}=2z_j \cos\gamma^{(l)}_j(\kappa)
\label{eq:linear2}
\end{eqnarray}
where $\gamma^{(l)}_j(\kappa)$ denotes  
\begin{eqnarray}
\gamma^{(l)}_{j}(\kappa)=
\left\{
\begin{array}{ll}
(2 \cdot 3^{j} -2^{j-l} 3^l)\kappa 
- 2^{j-l}(3^{l-1}\kappa_l^*+\pi+O(z_l)),
\quad 
& \mbox{for $l\neq 0$}\\
(2 \cdot 3^{j} -2^{j})\kappa 
- 2^{j}(m\pi+O(z_0)),
& \mbox{for $l=0$}
\end{array} 
\right.
\quad (\mbox{mod $\pi$}).
\label{phi}
\end{eqnarray}
Therefore, for $\kappa$ near $\kappa_l^*$, $z_j$ is given by
\begin{eqnarray}
z_j= z_l \prod_{i=l}^{j-1} \left(
2 \cos \gamma^{(l)}_i(\kappa)
\right).
\label{zj_expression}
\end{eqnarray}
Using this, we obtain the following transmission
coefficient $T_j$ near $\kappa^*_l$,
\begin{eqnarray}
T_j=1-z_l^2 \prod_{i=l}^{j-1} (2\cos \gamma^{(l)}_{i}(\kappa))^2.
\label{scale1_2}
\end{eqnarray}

As is proved in Appendix \ref{sec:appendix}, we can show that if
$\kappa_l^*/\pi$ is a rational number $s/t$ with coprime integers $s$
and $t$, the phase
$\gamma_i^{(l)}(\kappa_l^*)$ becomes periodic
for a sufficiently large $i$,
\begin{eqnarray}
\gamma_{i+p}^{(l)}(\kappa_l^*)=\gamma_i^{(l)}(\kappa_l^*),
\quad (i\ge r; \mbox{mod $\pi$}), 
\label{phi_cond}
\end{eqnarray}
where the minimal period $p$ is given by a divisor of $\varphi(t)$. (Here
$\varphi(x)$ is the Euler's totient function\cite{Euler}.)
Using this, we obtain a local scaling behavior near complete
transmission.
To see this, rewrite Eq.(\ref{scale1_2}) by using 
$\kappa=\kappa_l^*+\theta$ ($\theta\ll 1$),
\begin{eqnarray}
T_j(\theta)=1-z_l^2\prod_{i=l}^{j-1}
\left(2\cos [\gamma_i^{(l)}(\kappa_l^*)
+\gamma_i^{(l)'}(\kappa_l^*)\theta]\right)^2. 
\label{eq:scale2}
\end{eqnarray}
If $\kappa_l^*/\pi$ is a rational number $s/t$, then from
the relation (\ref{phi_cond}),
the product in the right-hand side of (\ref{eq:scale2}) 
for $j=pn+m-1$ ($m=1,2,\cdots, p$) is rewritten as
\begin{eqnarray}
&&\prod_{i=l}^{pn+m-2} 
\left(2\cos[\gamma^{(l)}_{i}(\kappa_l^*)
+\gamma_{i}^{(l)'}(\kappa_l^*)\theta]\right)^2
\nonumber\\
&&=\prod_{i=p(n-1)+m-1}^{pn+m-2} 
\left(2\cos[\gamma^{(l)}_{i}(\kappa_l^*)
+\gamma_{i}^{(l)'}(\kappa_l^*)\theta]\right)^2
\prod_{i=l}^{p(n-1)+m-2}
\left(2\cos[\gamma^{(l)}_{i}(\kappa_l^*)
+\gamma_{i}^{(l)'}(\kappa_l^*)\theta]\right)^2
\nonumber\\
&&=\prod_{\alpha=0}^{p-1} 
\left(2\cos[\gamma^{(l)}_{p(n-1)+m-1+\alpha}(\kappa_l^*)
+\gamma_{p(n-1)+m-1+\alpha}^{(l)'}(\kappa_l^*)\theta]\right)^2
\prod_{i=l}^{p(n-1)+m-2}
\left(2\cos[\gamma^{(l)}_{i}(\kappa_l^*)
+\gamma_{i}^{(l)'}(\kappa_l^*)\theta]\right)^2
\nonumber\\
&&=\prod_{\alpha=0}^{p-1} 
\left(2\cos[\gamma^{(l)}_{pn_0(m)+m-1+\alpha}(\kappa_l^*)
+\gamma_{p(n-1)+m-1+\alpha}^{(l)'}(\kappa_l^*)\theta]\right)^2
\prod_{i=l}^{p(n-1)+m-2}
\left(2\cos[\gamma^{(l)}_{i}(\kappa_l^*)
+\gamma_{i}^{(l)'}(\kappa_l^*)\theta]\right)^2,
\nonumber\\
\label{scale2_pre}
\end{eqnarray}
where $n_0(m)$ is the minimal integer satisfying $pn_0+m-1\ge r$. 
Therefore, we have
\begin{eqnarray}
1-T_{pn+m-1}(\theta)=
\prod_{\alpha=0}^{p-1} 
\left(2\cos[\gamma^{(l)}_{pn_0(m)+m-1+\alpha}(\kappa_l^*)
+\gamma_{p(n-1)+m-1+\alpha}^{(l)'}(\kappa_l^*)\theta]\right)^2
\left(
1-T_{p(n-1)+m-1}(\theta)
\right).
\nonumber\\
\end{eqnarray}
From Eq. (\ref{phi}), 
$\gamma_{p(n-1)+m-1+\alpha}^{(l)'}(\kappa_l^*)$ behaves as
\begin{eqnarray}
\gamma_{p(n-1)+m-1+\alpha}^{(l)'}(\kappa_l^*)
\sim
2\cdot 3^{p(n-1)+m-1+\alpha},
\end{eqnarray}
for $n\gg 1$.
Thus defining $f^{(m)}(\theta)$ as
\begin{eqnarray}
f^{(m)}(\theta)= 
\prod_{\alpha=0}^{p-1} 
\left(2\cos[\gamma^{(l)}_{pn_0(m)+m-1+\alpha}(\kappa_l^*)
+2\cdot 3^{\alpha}\theta]\right)^2,
\label{eq:scalingfunction}
\end{eqnarray}
we obtain
\begin{eqnarray}
1-T_{pn+m-1}(\theta)=f^{(m)}(3^{p(n-1)+m-1}\theta)
\left(1-T_{p(n-1)+m-1}(\theta)
\right)
\quad (n\gg 1).
\label{eq:TT}
\end{eqnarray}
By renormalizing $T_{pn+m-1}(\theta)$ as
\begin{eqnarray}
\hat{T}_{pn+m-1}(\theta)\equiv T_{pn+m-1}(\theta/3^{p(n-1)+m-1}), 
\end{eqnarray}
Eq.(\ref{eq:TT}) is rewritten as
\begin{eqnarray}
1-\hat{T}_{pn+m-1}(\theta)=f^{(m)}(\theta)(1-\hat{T}_{p(n-1)+m-1}(\theta/3^p)).
\label{eq:localscaling}
\end{eqnarray}
This equation clearly indicates that the ratio
$(1-\hat{T}_{pn+m-1}(\theta))/(1-\hat{T}_{p(n-1)+m-1}(\theta/3^p))$ does not
depend on the generation of Cantor sequences and it has a
scaling behavior with the scaling function $f^{(m)}(\theta)$.

 \begin{figure}
  \begin{center}
   \includegraphics[width=7.0cm,clip]{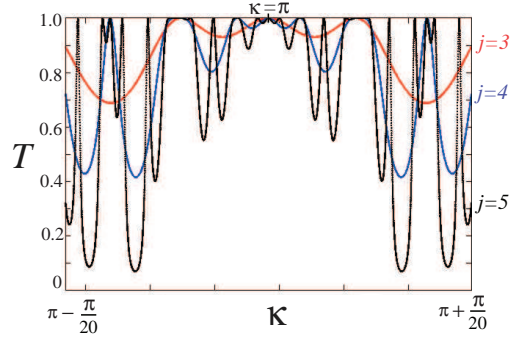}
  \end{center}
  \caption{(Color online) Transmission coefficients $T_j$ as
  functions of $\kappa$ for $n_{\rm A}=2.0$ around $\kappa_0^*=\pi$.
The generations are $j=3$, $j=4$, and $j=5$.}
  \label{fig:scaling_1}
 \end{figure}
  \begin{figure}
  \begin{center}
\includegraphics[width=8.5cm,clip]{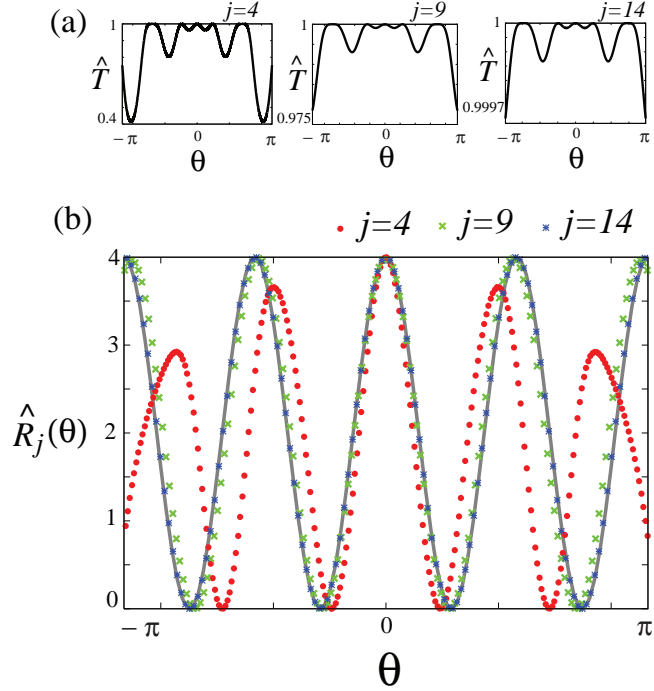}
 \caption{(Color online)
(a) Transmission coefficients $\hat{T}_j$ as
  functions of $\theta$ for $n_{\rm A}=2.0$ around $\kappa_0^*=\pi$.
 The generations are $j=4$, $j=9$, and $j=14$.
(b) The ratio $\hat{R}_j(\theta)=(1-\hat{T}_{j}(\theta))/(1-\hat{T}_{j-1}(\theta/3))$ 
for $n_{\rm A}=2.0$ around $\kappa_0^*=\pi$.
The generations are $j=4$, $j=9$, and $j=14$.
The scaling function $f^{(1)}(\theta)$ is also plotted by the line.}
\label{fig:scaling1_2}
\end{center}
 \end{figure}

To illustrate the local scaling behavior of $T_j$ near $T=1$ obtained above,
we compare our formula (\ref{eq:localscaling}) with numerical results
for various $\kappa_l^*$.
In Figs.\ref{fig:scaling_1} and \ref{fig:scaling1_2}, we show the transmission
coefficient $T_j$ for $n_{\rm A}=2$ and $\kappa_0^*=\pi$.
The transmission coefficients $T_j$ as functions of $\kappa$ are presented in
Fig.\ref{fig:scaling_1} and the renormalized one
$\hat{T}_j(\theta)$ is in Fig.\ref{fig:scaling1_2}.
In this case, the period $p$ in (\ref{phi_cond}) is $p=\varphi(1)=1$,
and the scaling function $f^{(1)}(\theta)$ is
\begin{eqnarray}
f^{(1)}(\theta)=4\cos^2(2\theta). 
\end{eqnarray}
For $j\ge 9$ we find an excellent agreement between our formula
(\ref{eq:localscaling}) and numerical data in
Fig.\ref{fig:scaling1_2}.

\begin{figure}
\begin{center}
\includegraphics[width=8.5cm,clip]{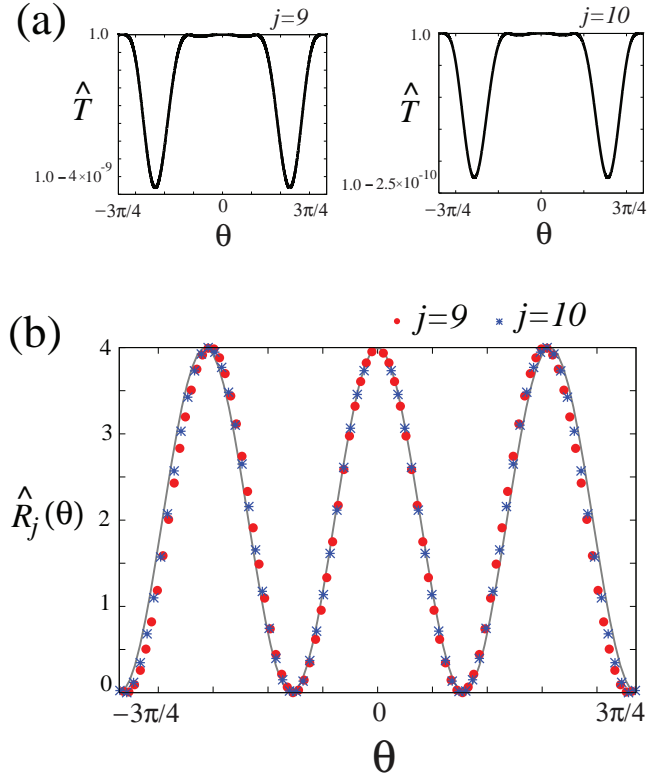}
 \caption{(Color online)
(a) Transmission coefficients $\hat{T}_j$ as
  functions of $\theta$ for $n_{\rm A}=2.0$ around $\kappa_0^*=\pi/2$.
  The generations are $j=9$ and $j=10$.
(b) The ratio $\hat{R}_j(\theta)=(1-\hat{T}_{j}(\theta))/(1-\hat{T}_{j-1}(\theta/3))$ 
for $n_{\rm A}=2.0$ around $\kappa_0^*=\pi/2$.
The generations are $j=9$ and $j=10$.
The scaling function $f^{(1)}(\theta)$ is also plotted by the line.}
\label{fig:linear_2}
\end{center}
\end{figure}

\begin{figure}
\begin{center}
\includegraphics[width=8.0cm,clip]{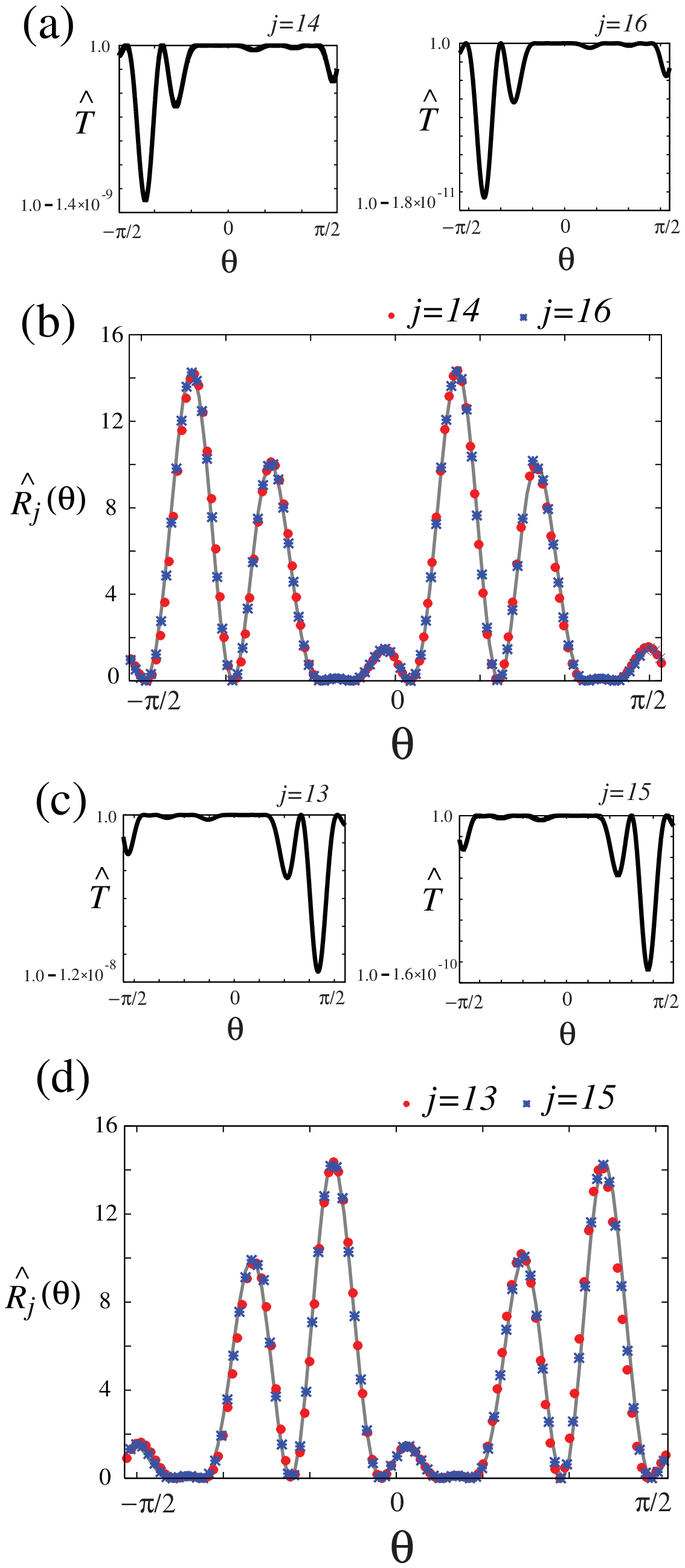}
 \caption{(Color online)
(a) Transmission coefficients $\hat{T}_j$ as
  functions of $\theta$ for $n_{\rm A}=3.0$ around $\kappa_0^*=\pi/3$.
  The generations are $j=14$ and $j=16$.
(b) The ratio $\hat{R}_j(\theta)=(1-\hat{T}_{j}(\theta))/(1-\hat{T}_{j-2}(\theta/3^2))$ 
for $n_{\rm A}=3.0$ around $\kappa_0^*=\pi/3$.
The generations are $j=14$ and $j=16$.
The scaling function $f^{(1)}(\theta)$ is also plotted by the line.
(c) $\hat{T}_j$ for the generations $j=13$ and $j=15$.
(d) $\hat{R}_j$ for the generations $j=13$ and $j=15$.
The scaling function is $f^{(2)}(\theta)$ in this case.}
\label{fig:linear_4}
\end{center}
\end{figure}

\begin{figure}
\begin{center}
\includegraphics[width=8.0cm,clip]{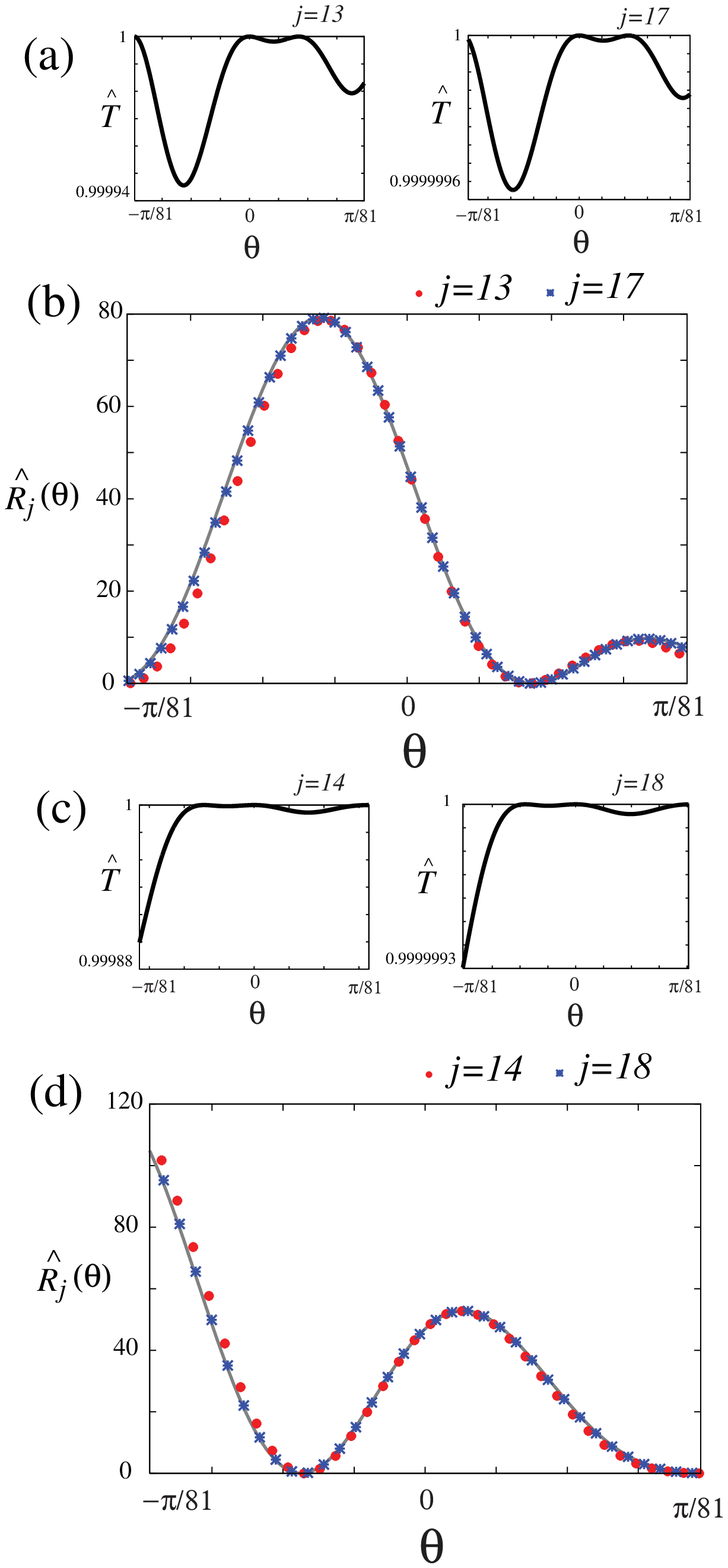}
 \caption{(Color online)
(a) Transmission coefficients $\hat{T}_j$ as
  functions of $\theta$ for $n_{\rm A}=5.0$ around $\kappa_0^*=\pi/5$.
  The generations are $j=13$ and $j=17$.
(b) The ratio $\hat{R}_j(\theta)=(1-\hat{T}_{j}(\theta))/(1-\hat{T}_{j-4}(\theta/3^4))$ 
for $n_{\rm A}=5.0$ around $\kappa_0^*=\pi/5$.
The generations are $j=13$ and $j=17$.
The scaling function $f^{(2)}(\theta)$ is also plotted by the line.
(c) $\hat{T}_j$ for the generations $j=14$ and $j=18$.
(d) $\hat{R}_j$ for the generations $j=14$ and $j=18$.
The scaling function is $f^{(3)}(\theta)$ in this case.}
\label{fig:linear_5}
\end{center}
\end{figure}

In Figs.\ref{fig:linear_2}-\ref{fig:linear_5}, we also present local scaling
behaviors for $n_{\rm A}=t$ and $\kappa_0^*=\pi/t$ with integers $t=2,3,5$.
For $t=2,3,5$, the period $p$ is given by $p=1,2,4$, respectively.  
The scaling functions $f^{(m)}(\theta)$ are given by
\begin{eqnarray}
f^{(1)}(\theta)=4\cos^2(2\theta),
\end{eqnarray}
for $t=2$, and
\begin{eqnarray}
&&f^{(1)}(\theta)=16\cos^2(\pi/3+6\theta)\cos^2(2\pi/3+2\theta),
\nonumber\\
&&f^{(2)}(\theta)=16\cos^2(2\pi/3+6\theta)\cos^2(\pi/3+2\theta), 
\end{eqnarray}
for $t=3$, and
\begin{eqnarray}
&&f^{(1)}(\theta)=256\cos^2\left(\pi/5+54\theta\right)
\cos^2\left(4\pi/5+18\theta\right)\cos^2\left(4\pi/5+6\theta\right)
\cos^2\left(\pi/5+2\theta\right),
\nonumber\\
&&f^{(2)}(\theta)=256\cos^2\left(\pi/5+54\theta\right)
\cos^2\left(\pi/5+18\theta\right)
\cos^2\left(4\pi/5+6\theta\right)\cos^2\left(4\pi/5+2\theta\right),
\nonumber\\
&&f^{(3)}(\theta)=256\cos^2\left(4\pi/5+54\theta\right)
\cos^2\left(\pi/5+18\theta\right)\cos^2\left(\pi/5+6\theta\right)
\cos^2\left(4\pi/5+2\theta\right),
\nonumber\\
&&f^{(4)}(\theta)=256\cos^2\left(4\pi/5+54\theta\right)
\cos^2\left(4\pi/5+18\theta\right)\cos^2\left(\pi/5+6\theta\right)
\cos^2\left(\pi/5+2\theta\right),
\end{eqnarray}
for $t=5$.
Again, we have excellent agreements between the numerical data and 
our analytical results.

\begin{figure}
\begin{center}
\includegraphics[width=8.5cm,clip]{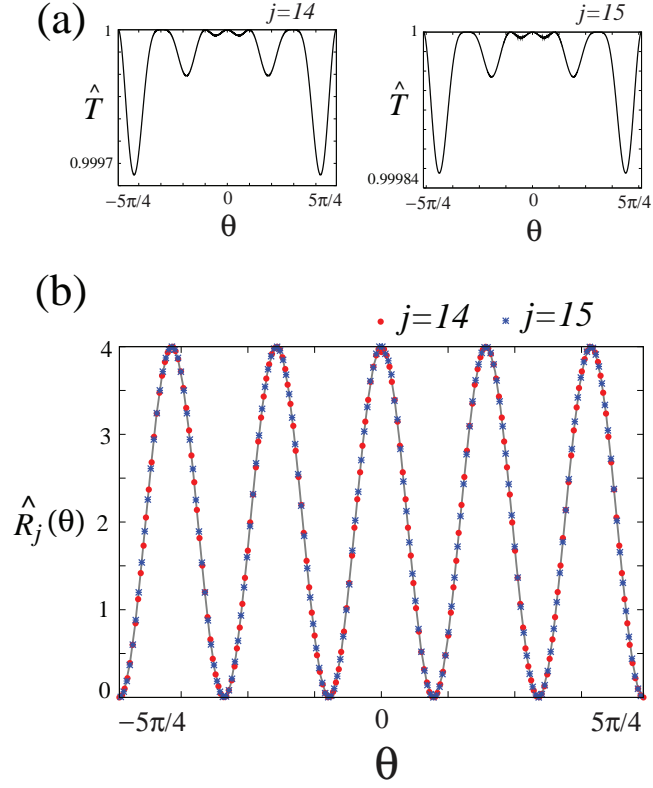}
 \caption{(Color online)
(a) Transmission coefficients $T_j$ as
  functions of $\theta$ for $n_{\rm A}=1/2$ around $\kappa_1^*=\pi$.
 The generations are $j=14$ and $j=15$.
(b) The ratio $\hat{R}_j(\theta)=(1-\hat{T}_{j}(\theta))/(1-\hat{T}_{j-1}(\theta/3))$ 
for $n_{\rm A}=1/2$ around $\kappa_1^*=\pi$.
The generations are $j=14$ and $j=15$.
The scaling function $f^{(1)}(\theta)$ is also plotted by the line.}
\label{fig:linear_3}
 \end{center}
 \end{figure}

As an example with $l\neq 0$, we show a local scaling behavior for
$n_{\rm A}=1/2$ and $\kappa_1^{*}=\pi$ in Fig.\ref{fig:linear_3}.
Here $p=1$ and the scaling function is given by
\begin{eqnarray}
f^{(1)}(\theta)=4\cos^2(2\theta). 
\label{scaling_T_3}
\end{eqnarray}
It also reproduces the numerical data excellently. 

\begin{figure}
\begin{center}
    \includegraphics[width=8.5cm,clip]{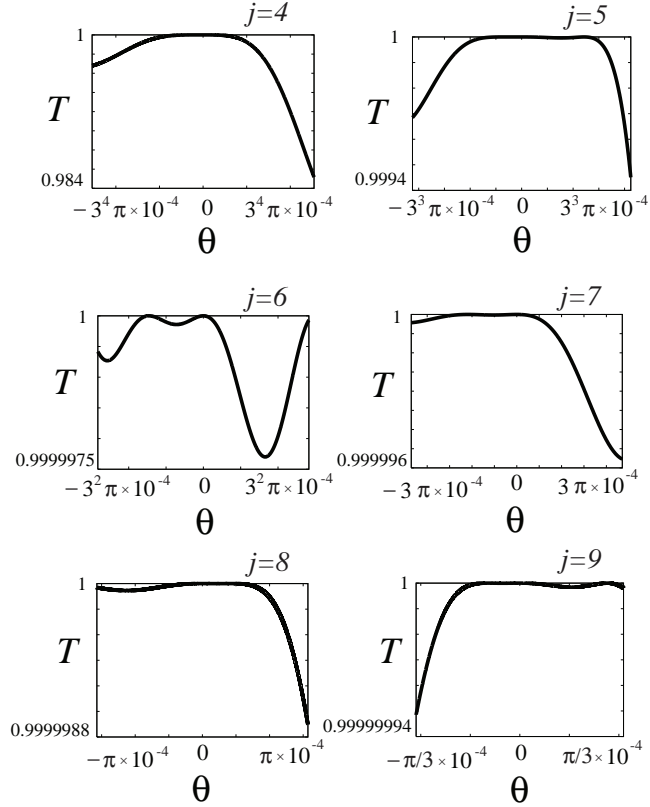}
\end{center}
   \caption{Transmission coefficients $T_j$ as
  functions of $\theta$ for $n_{\rm A}=\sqrt2$ around 
$\kappa_0^*=\frac{\sqrt{2}}{2}\pi$.
The generations are from $j=4$ to $j=9$.
 The horizontal axes are rescaled by a factor of three
as a generation increases.}
   \label{fig:noscale}
 \end{figure}

We also confirm numerically that if complete transmission is
achieved for a wave number with an irrational $\kappa^*/\pi$, no scaling
behavior is obtained.
The case of $n_{\rm A}=\sqrt{2}$ is shown in Fig.\ref{fig:noscale}.
Although complete transmission is realized at
$\kappa=\frac{\sqrt{2}}{2}\pi$,
no scaling behavior of $T_j$ near the complete transmission is found in 
Fig.\ref{fig:noscale}.

\section{Summary and discussions}
\label{sec:summary}

We investigate wave propagation through Cantor-set media 
on the basis of the renormalization-group equation. 
We analytically find that, for specific wave numbers,
the transmission coefficients are governed by the logistic map.
Especially, in the chaotic region, the transmission coefficients  
show sensitive dependence on small changes of parameters of the system
such as the index of refraction.
For wave numbers near the values giving the chaotic behavior,
the transmission coefficients again show chaotic behaviors
in the first few generations.
 For other values of wave numbers, our numerical results suggest that 
light transmits completely or reflects completely by the Cantor-set media.
We also show that the transmission coefficients exhibit local scaling behaviors
near complete transmission if the complete transmission is 
achieved at a wave number $\kappa^*$ with a rational $\kappa^*/\pi$.
The analytic form of the scaling function is determined by $\kappa^*$
through the Euler's totient function.

Finally, we would like to mention that
our approach developed here can be extended to the generalized
Cantor sequences, where the length of each A becomes $1/\beta$
as one increases the generation.
For a positive integer $\beta$, in a similar manner as Sec.\ref{sec:chaos}, 
it is found that the initial wave numbers which give chaotic behaviors are
\begin{eqnarray}
\kappa=\frac{2 m}{(\beta-1)(\beta-2)\beta^{q}}\pi,
\quad (m,q:{\rm integers}).
\label{coherentk_general}
\end{eqnarray}
In addition, for an odd positive integer $\beta$,
chaotic behaviors are also found to appear for
\begin{eqnarray}
\kappa=\frac{2m+1}{(\beta-1)(\beta-2)\beta^{q}}\pi,
\quad (m,q:{\rm integers}).
\label{coherent2_general}
\end{eqnarray}

\appendix
\section{periodicity of $\gamma_i^{(l)}(\kappa_l^*)$}
\label{sec:appendix}

In this Appendix, we prove that $\gamma_i^{(l)}(\kappa_l^*)$ has the periodicity
(\ref{phi_cond}) if and only if $\kappa_l^{*}/\pi$ is a rational number. 
From the explicit forms of $\gamma_i^{(l)}$ given by (\ref{phi}), it is
immediately found that if $\gamma_i^{(l)}(\kappa_l^*)$ has the
periodicity (\ref{phi_cond}), then
$\kappa_l^*/\pi$ should be a rational number.
Therefore, we will show in the following that if $\kappa_l^{*}/\pi$ is a
rational number, then the periodicity (\ref{phi_cond}) is obtained.

To prove this, 
we use the Euler's theorem:
\begin{eqnarray}
N^{\varphi(M)}=1,
\quad(\mbox{mod}\, M),
\label{Euler}
\end{eqnarray}
where $N$ and $M$ are mutually prime integers, and 
$\varphi(M)$ is the Euler's totient function
 which counts the number of positive
integers not greater than and coprime to $M$ \cite{Euler}.
The Euler's totient function satisfies
\begin{eqnarray}
\varphi(nm)=\varphi(n)\varphi(m),
\label{multiple}
\end{eqnarray}
for coprime positive integers $n$ and $m$.

Since $O(z_l)$ corrections in Eq. (\ref{phi}) disappear for
$\kappa=\kappa_l^{*}$,
$\gamma_{i}^{(l)}(\kappa_l^*)$ is given by
\begin{eqnarray}
\gamma^{(l)}_{i}(\kappa_l^*)
=\left\{
\begin{array}{ll}
(2\cdot 3^{i}-2^{i-l+2}\cdot 3^{l-1})\kappa_l^*,
& \mbox{for $l\neq 0$}\\
(2\cdot 3^{i}-2^{i})\kappa_0^*,
& \mbox{for $l=0$}
\end{array} 
\quad(\mbox{mod}\, \pi).
\right.
\label{phi_1}
\end{eqnarray}
Thus for a rational $\kappa_l^*/\pi=s/ t$ with coprime integers $s$ and
$t$, we have
\begin{eqnarray}
\gamma^{(l)}_{i}(\kappa_l^*)
=\left\{
\begin{array}{ll}
\left(\frac{2\cdot 3^{i}}{t}-\frac{2^{i-l+2}\cdot 3^{l-1}}{t}\right)s\pi,
&\mbox{for $l\neq 0$}\\
\left(\frac{2\cdot 3^{i}}{t}-\frac{2^{i}}{t}\right)s\pi,
& \mbox{for $l=0$}
\end{array}
\right.
\quad(\mbox{mod}\, \pi).
\label{phi_2}
\end{eqnarray}
Now decompose $t$ into $t=2^{\eta} 3^{\xi} u$ where $\eta$, $\xi$ and
$u$ are integers and $u$ is coprime to $2$ and $3$.
Then we obtain
\begin{eqnarray}
\gamma^{(l)}_{i}(\kappa_l^*)
=\left\{
\begin{array}{ll}
\left(\frac{3^{i-\xi}}{2^{\eta-1} u}
-\frac{2^{i-l+2-\eta}}{3^{\xi-l+1} u}\right)s\pi,
& \mbox{for $l\neq 0$}\\
\left(\frac{3^{i-\xi}}{2^{\eta-1} u}
-\frac{2^{i-\eta}}{3^{\xi} u}\right)s\pi,
& \mbox{for $l=0$}
\end{array}
\quad(\mbox{mod}\, \pi).
\right.
\label{phi_3}
\end{eqnarray}
Since $3$ and $2^{\eta} u$, and $2$ and $3^{\xi} u$ 
are mutually prime integers, respectively, 
we have from the Euler's theorem
\begin{eqnarray}
3^{\varphi(2^{\eta}u)}&=&1,\quad(\mbox{mod}\, 2^{\eta} u),\nonumber\\
2^{\varphi(3^{\xi} u)}&=&1,\quad(\mbox{mod}\, 3^{\xi} u).
\label{prime_cycle}
\end{eqnarray}
Moreover, using the relation (\ref{multiple}), we find 
\begin{eqnarray}
3^{\varphi(2^{\eta} 3^{\xi}u)}
&=&(3^{\varphi(2^{\eta}u)})^{\varphi(3^{\xi})}=1,
\quad(\mbox{mod}\, 2^{\eta} u),\nonumber\\
2^{\varphi(2^{\eta} 3^{\xi} u)}
&=&(2^{\varphi(3^{\xi} u)})^{\varphi(2^{\eta})}=1,
\quad(\mbox{mod}\, 3^{\xi} u),
\label{prime_cycle2}
\end{eqnarray}
namely
\begin{eqnarray}
3^{\varphi(t)}=1+2^{\eta}u M,
\quad 
2^{\varphi(t)}=1+3^{\xi}u N,
\label{prime_cycle3}
\end{eqnarray}
with integers $M$ and $N$.
From (\ref{prime_cycle3}), we have
\begin{eqnarray}
\gamma_{i+\varphi(t)}^{(l)}(\kappa_l^*)
&=&
\left(
\frac{3^{i+\varphi(t)-\xi}}{2^{\eta-1}u}
-\frac{2^{i+\varphi(t)-l+2-\eta}}{3^{\xi-l+1}u}
\right)s\pi
\nonumber\\
&=&\left(
\frac{3^{i-\xi}}{2^{\eta-1}u}
-\frac{2^{i-l+2-\eta}}{3^{\xi-l+1}u}
\right)s\pi+(2\cdot3^{i-\xi}M-2^{i-l+2-\eta}3^{l-1}N)s\pi
\nonumber\\
&=&\gamma_i^{(l)}(\kappa_l^*) 
+(2\cdot3^{i-\xi}M-2^{i-l+2-\eta}3^{l-1}N)s\pi,
\end{eqnarray}
for $l\neq 0$, and  
\begin{eqnarray}
\gamma_{i+\varphi(t)}^{(l)}(\kappa_l^*)
&=&
\left(
\frac{3^{i+\varphi(t)-\xi}}{2^{\eta-1}u}
-\frac{2^{i+\varphi(t)-\eta}}{3^{\xi}u}
\right)s\pi
\nonumber\\
&=&\left(
\frac{3^{i-\xi}}{2^{\eta-1}u}
-\frac{2^{i-\eta}}{3^{\xi}u}
\right)s\pi+(2\cdot3^{i-\xi}M-2^{i-\eta}N)s\pi
\nonumber\\
&=&\gamma_i^{(l)}(\kappa_l^*) 
+(2\cdot3^{i-\xi}M-2^{i-\eta}N)s\pi,
\end{eqnarray}
for $l=0$.
Therefore, we obtain 
Eq.(\ref{phi_cond}) with $p=\varphi(t)$ for a sufficiently large $i$.
Here note that $\varphi(t)$ is not the minimal period of
$\gamma_i^{(l)}(\kappa_l^{*})$ in general. Thus the minimal period $p$ 
is a divisor of $\varphi(t)$.

\begin{acknowledgments}

We are grateful to M. Yamanaka for useful discussions.
This work was supported in part by Global COE Program
``the Physical Sciences Frontier'', MEXT, Japan for K.E.

\end{acknowledgments}

\end{document}